\input harvmac

\font\tenmsb=msbm10
\font\eighteenmsb=msbm18

\newfam\bigRfam
\textfont\bigRfam=\eighteenmsb

\newfam\msbfam
\textfont\msbfam=\tenmsb

\def\bigR{{\fam\bigRfam{R}}}
\def\A{{\cal{A}}}
\def\N{{\cal{N}}}
\def\R{{\fam\msbfam{R}}}
\def\V{{\cal{V}}}
\def\Z{{\cal{Z}}}
\def\ZZ{{\fam\msbfam{Z}}}

\font\bigcmsy=cmsy10 scaled 1500
\def\Bsw{\!\mathrel{\hbox{\bigcmsy\char'056}}\!}
\def\Bse{\!\mathrel{\hbox{\bigcmsy\char'046}}\!}

\lref\HawkingDH{
S.~W.~Hawking and D.~N.~Page,
``Thermodynamics of black holes in anti-de Sitter space,''
Commun.\ Math.\ Phys.\  {\bf 87}, 577 (1983).
}

\lref\WittenZW{
E.~Witten,
``Anti-de Sitter space, thermal phase transition, and confinement in  gauge
theories,''
Adv.\ Theor.\ Math.\ Phys.\  {\bf 2}, 505 (1998)
[arXiv:hep-th/9803131].
}

\lref\SundborgUE{
B.~Sundborg,
``The Hagedorn transition, deconfinement and $\N = 4$ SYM theory,''
Nucl.\ Phys.\ B {\bf 573}, 349 (2000)
[arXiv:hep-th/9908001].
}

\lref\EricksonAF{
J.~K.~Erickson, G.~W.~Semenoff and K.~Zarembo,
``Wilson loops in $\N = 4$ supersymmetric Yang-Mills theory,''
Nucl.\ Phys.\ B {\bf 582}, 155 (2000)
[arXiv:hep-th/0003055].
}

\lref\DrukkerRR{
N.~Drukker and D.~J.~Gross,
``An exact prediction of $\N = 4$ SUSYM theory for string theory,''
J.\ Math.\ Phys.\  {\bf 42}, 2896 (2001)
[arXiv:hep-th/0010274].
}

\lref\SundborgWP{
B.~Sundborg,
``Stringy gravity, interacting tensionless strings and massless higher
spins,''
Nucl.\ Phys.\ Proc.\ Suppl.\  {\bf 102}, 113 (2001)
[arXiv:hep-th/0103247].
}

\lref\PolyakovAF{
A.~M.~Polyakov,
``Gauge fields and space-time,''
Int.\ J.\ Mod.\ Phys.\ A {\bf 17S1}, 119 (2002)
[arXiv:hep-th/0110196].
}

\lref\BerensteinJQ{
D.~Berenstein, J.~M.~Maldacena and H.~Nastase,
``Strings in flat space and pp waves from $\N = 4$ super Yang Mills,''
JHEP {\bf 0204}, 013 (2002)
[arXiv:hep-th/0202021].
}

\lref\GubserTV{
S.~S.~Gubser, I.~R.~Klebanov and A.~M.~Polyakov,
``A semi-classical limit of the gauge/string correspondence,''
Nucl.\ Phys.\ B {\bf 636}, 99 (2002)
[arXiv:hep-th/0204051].
}

\lref\ConstableHW{
N.~R.~Constable, D.~Z.~Freedman, M.~Headrick, S.~Minwalla, L.~Motl,
A.~Postnikov and W.~Skiba,
``PP-wave string interactions from perturbative Yang-Mills theory,''
JHEP {\bf 0207}, 017 (2002)
[arXiv:hep-th/0205089].
}

\lref\SantambrogioSB{
A.~Santambrogio and D.~Zanon,
``Exact anomalous dimensions of $\N = 4$ Yang-Mills operators with large R
charge,''
Phys.\ Lett.\ B {\bf 545}, 425 (2002)
[arXiv:hep-th/0206079].
}

\lref\KlebanovMP{
I.~R.~Klebanov, M.~Spradlin and A.~Volovich,
``New effects in gauge theory from pp-wave superstrings,''
Phys.\ Lett.\ B {\bf 548}, 111 (2002)
[arXiv:hep-th/0206221].
}

\lref\HalpernFZ{
M.~B.~Halpern,
``On the large $N$ limit of conformal field theory,''
Annals Phys.\  {\bf 303}, 321 (2003)
[arXiv:hep-th/0208150].
}

\lref\BeisertBB{
N.~Beisert, C.~Kristjansen, J.~Plefka, G.~W.~Semenoff and M.~Staudacher,
``BMN correlators and operator mixing in $\N = 4$ super Yang-Mills theory,''
Nucl.\ Phys.\ B {\bf 650}, 125 (2003)
[arXiv:hep-th/0208178].
}

\lref\GrossMH{
D.~J.~Gross, A.~Mikhailov and R.~Roiban,
``A calculation of the plane wave string Hamiltonian from $\N = 4$
super-Yang-Mills theory,''
JHEP {\bf 0305}, 025 (2003)
[arXiv:hep-th/0208231].
}

\lref\ConstableVQ{
N.~R.~Constable, D.~Z.~Freedman, M.~Headrick and S.~Minwalla,
``Operator mixing and the BMN correspondence,''
JHEP {\bf 0210}, 068 (2002)
[arXiv:hep-th/0209002].
}

\lref\DolanZH{
F.~A.~Dolan and H.~Osborn,
``On short and semi-short representations for four dimensional superconformal
symmetry,''
Annals Phys.\  {\bf 307}, 41 (2003)
[arXiv:hep-th/0209056].
}

\lref\PearsonZS{
J.~Pearson, M.~Spradlin, D.~Vaman, H.~Verlinde and A.~Volovich,
``Tracing the string: BMN correspondence at finite $J^2/N$,''
JHEP {\bf 0305}, 022 (2003)
[arXiv:hep-th/0210102].
}

\lref\MinahanVE{
J.~A.~Minahan and K.~Zarembo,
``The Bethe-ansatz for $\N = 4$ super Yang-Mills,''
JHEP {\bf 0303}, 013 (2003)
[arXiv:hep-th/0212208].
}

\lref\BeisertFF{
N.~Beisert, C.~Kristjansen, J.~Plefka and M.~Staudacher,
``BMN gauge theory as a quantum mechanical system,''
Phys.\ Lett.\ B {\bf 558}, 229 (2003)
[arXiv:hep-th/0212269].
}

\lref\BeisertTQ{
N.~Beisert, C.~Kristjansen and M.~Staudacher,
``The dilatation operator of $\N = 4$ super Yang-Mills theory,''
Nucl.\ Phys.\ B {\bf 664}, 131 (2003)
[arXiv:hep-th/0303060].
}

\lref\BianchiWX{
M.~Bianchi, J.~F.~Morales and H.~Samtleben,
``On stringy AdS${}_5 \times S^5$ and higher spin holography,''
JHEP {\bf 0307}, 062 (2003)
[arXiv:hep-th/0305052].
}

\lref\BenaWD{
I.~Bena, J.~Polchinski and R.~Roiban,
``Hidden symmetries of the $AdS_5 \times S^5$ superstring,''
Phys.\ Rev.\ D {\bf 69}, 046002 (2004)
[arXiv:hep-th/0305116].
}

\lref\BeisertJJ{
N.~Beisert,
``The complete one-loop dilatation operator of $\N = 4$ super Yang-Mills
theory,''
Nucl.\ Phys.\ B {\bf 676}, 3 (2004)
[arXiv:hep-th/0307015].
}

\lref\BeisertYB{
N.~Beisert and M.~Staudacher,
``The $\N = 4$ SYM integrable super spin chain,''
Nucl.\ Phys.\ B {\bf 670}, 439 (2003)
[arXiv:hep-th/0307042].
}

\lref\DolanUH{
L.~Dolan, C.~R.~Nappi and E.~Witten,
``A relation between approaches to integrability in superconformal Yang-Mills
theory,''
JHEP {\bf 0310}, 017 (2003)
[arXiv:hep-th/0308089].
}

\lref\AharonySX{
O.~Aharony, J.~Marsano, S.~Minwalla, K.~Papadodimas and M.~Van Raamsdonk,
``The Hagedorn/deconfinement phase transition in weakly coupled large $N$ gauge
theories,''
arXiv:hep-th/0310285.
}

\lref\TseytlinII{
A.~A.~Tseytlin,
``Spinning strings and AdS/CFT duality,''
arXiv:hep-th/0311139.
}

\lref\DolanPS{
L.~Dolan, C.~R.~Nappi and E.~Witten,
``Yangian symmetry in D = 4 superconformal Yang-Mills theory,''
arXiv:hep-th/0401243.
}

\lref\RyzhovNZ{
A.~V.~Ryzhov and A.~A.~Tseytlin,
``Towards the exact dilatation operator of $\N = 4$ super Yang-Mills theory,''
arXiv:hep-th/0404215.
}

\lref\BeisertDI{
N.~Beisert, M.~Bianchi, J.~F.~Morales and H.~Samtleben,
``Higher spin symmetry and $\N = 4$ SYM,''
JHEP {\bf 0407}, 058 (2004)
[arXiv:hep-th/0405057].
}

\lref\FidkowskiFC{
L.~Fidkowski and S.~Shenker,
``D-brane instability as a large N phase transition,''
arXiv:hep-th/0406086.
}

\lref\AharonyIG{
O.~Aharony, J.~Marsano, S.~Minwalla and T.~Wiseman,
``Black hole-black string phase transitions in thermal 1+1 dimensional
supersymmetric Yang-Mills theory on a circle,''
arXiv:hep-th/0406210.
}

\lref\ArutyunovVX{
G.~Arutyunov, S.~Frolov and M.~Staudacher,
``Bethe ansatz for quantum strings,''
arXiv:hep-th/0406256.
}

\lref\TseytlinCJ{
A.~A.~Tseytlin,
``Semiclassical strings in $AdS_5 \times S^5$ and scalar operators in
$\N = 4$ SYM
theory,''
arXiv:hep-th/0407218.
}

\lref\BeisertRY{
N.~Beisert,
``The dilatation operator of $\N = 4$
super Yang-Mills theory and integrability,''
arXiv:hep-th/0407277.
}

\lref\LiuVY{
H.~Liu,
``Fine structure of Hagedorn transitions,''
arXiv:hep-th/0408001.
}

\lref\MMP{
M.~Spradlin, M.~Van Raamsdonk and A.~Volovich,
``Two-Loop Partition Function in the Planar Plane-Wave Matrix Model,''
arXiv:hep-th/0409178.
}

\lref\ThreeLoop{
O.~Aharony, J.~Marsano, S.~Minwalla,
K.~Papadodimas and M.~Van Raamsdonk, work in progress.
}

\Title
{\vbox{
 \baselineskip12pt
\hbox{hep-th/0408178}
\hbox{NSF-KITP-04-99}
}}
{\vbox{
\centerline{A Pendant for P\'olya:}
\centerline{}
\centerline{The One-Loop Partition Function of
${\cal{N}} = 4$ SYM on $\bigR \times S^3$}
}}

\centerline{
Marcus Spradlin and Anastasia Volovich
}

\bigskip
\centerline{Kavli Institute for Theoretical Physics}
\centerline{University of California}
\centerline{Santa Barbara, CA 93106 USA}
\centerline{\tt spradlin, nastja@kitp.ucsb.edu}

\vskip .3in
\centerline{\bf Abstract}

We study weakly coupled $SU(N)$ $\N = 4$ super Yang-Mills theory on
$\R \times S^3$ at infinite~$N$, which has interesting thermodynamics,
including a Hagedorn transition, even at zero Yang-Mills coupling.
We calculate the exact one-loop partition function below
the Hagedorn temperature.  Our calculation employs the representation of the
one-loop dilatation operator as a spin chain Hamiltonian acting on neighboring
sites and a generalization of P\'olya's counting of `necklaces'
(gauge-invariant operators) to include necklaces with a `pendant' (an operator
which acts on neighboring beads).
We find that the one-loop correction to the Hagedorn temperature is
$\delta \ln T_{\rm H} = + \lambda/8 \pi^2$.

\Date{}

\listtoc
\writetoc

\newsec{Introduction}

The past several years have witnessed
a tremendous amount of effort 
invested into the careful study of
$\N = 4$ super Yang-Mills (SYM) theory at large $N$ from a number
of complementary approaches.
One motivation for much of this work is the fact that
this theory is believed to provide, via the AdS/CFT
correspondence, the simplest context in which
we might hope
to understand how to solve large $N$ gauge theories in four dimensions.
Optimistically, the apparent integrability \BenaWD\ of the 
string theory dual to $\N = 4$ SYM theory suggests that
it might be possible to calculate
all physical quantities
(at least at infinite $N$)
as exact functions of the 't Hooft parameter
$\lambda = g_{\rm YM}^2 N$.
The successful calculation of the
circular BPS Wilson loop 
for all $\lambda$
\refs{\EricksonAF,\DrukkerRR} provided an early realization of this hope.

More recently, significant progress has been made on the
problem of calculating anomalous dimensions of gauge theory operators.
Two related approaches relevant to this problem include
the study of semiclassical string solutions following
\GubserTV, and the
study of the dilatation operator directly as the Hamiltonian
of an integrable spin chain following \refs{\MinahanVE,\BeisertYB}.
Comprehensive reviews of the most recent progress on these
approaches can be found in \refs{\TseytlinII,\TseytlinCJ}
and \BeisertRY.
BMN operators provide a prime example of a class of operators
whose anomalous dimensions can be calculated for all $\lambda$
\refs{\BerensteinJQ,\SantambrogioSB}.
Very recent work with partial results on the problem of summing
up all orders in $\lambda$ includes
\refs{\RyzhovNZ,\ArutyunovVX}.

In this paper we continue in the fine tradition of chipping away
at $\N = 4$ SYM theory from a variety of angles.
Our present interest lies in the partition function of the
theory
on $\R \times S^3$ at infinite $N$, which
displays interesting thermodynamic behavior \WittenZW, including
a Hagedorn transition \SundborgUE, 
even at zero 't Hooft parameter.
The partition function can be calculated  exactly
at $\lambda = 0$
by simply counting gauge-invariant operators using P\'olya theory
(see for
example \refs{\SundborgUE,\SundborgWP,
\AharonySX,\BianchiWX}). Here we calculate
the one-loop correction to the partition
function below the Hagedorn temperature and find
the result
\eqn\mainresult{{
\left. \ln \Tr[e^{-\beta H}]\right|_{\rm 1-loop}
=- \beta{\lambda \over 4 \pi^2}
\sum_{k=1}^\infty
\left[ k { \langle  D_2(x_k)
\rangle
\over 1- z(x_k)
} + \sum_{m=1}^\infty \langle P D_2(x_k,x_m)
\rangle
\right],
}}
where $x_n = \omega^{n+1} e^{-\beta n}$,
$\omega = e^{2 \pi i}$ (so that $\omega^{m/2} = \pm 1$
depending on whether $m$ is even or odd),
and the function
$z(x)$ is given below in (2.4).
The remaining quantities $\langle D_2(x) \rangle$ and
$\langle P D_2(w,y) \rangle$
are traces of the one-loop dilatation operator $D_2$
acting on two neighboring
fields inside
an operator.  We evaluate these traces explicitly in (5.14) and (5.17) below.
We also obtain from \mainresult\ 
the
one-loop correction to the value
of the Hagedorn temperature,
\eqn\aaa{
T_{\rm H}(\lambda) = T_{{\rm H}}(0)
\left(1 + {\lambda \over 8 \pi^2} + \cdots\right),
\qquad T_{{\rm H}}{(0)} \approx 0.380
}
(measured in units where the radius of the $S^3$ is one),
which is presumably another example
of an interesting physical quantity which we might hope to one day
calculate as an exact function of $\lambda$.

We begin in section 2 by reviewing how the free partition function
may be calculated by first assembling the partition function for
the elementary
fields into a partition function for single-trace operators, and then
into the full partition function for multi-trace operators.
We discuss the general structure of the one-loop correction
to the partition function and comment on
the role of operator mixing and $1/N$ effects.
In section 3 we rephrase the problem of counting gauge theory
operators into the language of spin chains and introduce a generalization
of P\'olya's necklace problem to what we call necklaces with a `pendant'.
These are necklaces with some local operator
$\widehat{O}$, such as a spin chain
Hamiltonian, inserted at neighboring beads
on the necklace.  We derive a general formula
expressing a partition function for such necklaces in terms
of some basic quantities
$\langle O(x)\rangle$ and $\langle PO(w,y)\rangle$ which are easily
obtained from $\widehat{O}$.
In section 4 we apply this machinery to various familiar subsectors
($SU(2)$, $SO(6)$, and $SL(2)$)
of the $\N = 4$ SYM theory.  This section is mostly a warm-up for
section 5, where we calculate $\langle D_2(x)\rangle$
and $\langle PD_2(w,y)\rangle$ for the  one-loop dilatation
operator $D_2$ of the full $\N = 4$ SYM theory.
Finally in section 6 we present
our final results for the one-loop corrections to the single- and 
multi-trace partition functions of $\N=4$ SYM.

\newsec{The $\N = 4$ SYM Partition Function}

We begin in this section
with a discussion of weakly-coupled $\N = 4$
SYM theory on $\R \times S^3$
following \refs{\SundborgUE,
\AharonySX}.  
Since there is more than one way to calculate the free
partition function, we review here that derivation which best sets
the stage for our calculation of the one-loop correction in the
following sections.

\subsec{Initial considerations}

In general, the partition function is
given by
\eqn\zbeta{
{\cal{Z}}(\beta) = \Tr[e^{-\beta H}],
}
where $\beta$ is the inverse temperature and $H$ is the Hamiltonian
of the theory on $\R \times S^3$.
It is convenient to introduce the bookkeeping parameter
\eqn\defx{
x = e^{-\beta} = e^{-1/T},
}
which ranges from $x = 0$ (zero temperature) to $x = 1$ (infinite
temperature).
According to the state-operator map there is a one-to-one correspondence
between states of the theory on $\R \times S^3$ and gauge-invariant
operators on $\R^4$.  The Hamiltonian $H$ on $\R \times S^3$ is identified
with the dilatation operator $D$ on the plane,
so counting states weighted by $x$ to the power of
their energy is equivalent
to counting gauge-invariant operators weighted by
$x$ to the power of their dimension on the plane.
Therefore the partition function can be written as
\eqn\zx{
\Z(x) = \Tr[x^{D}],
}
where we implicitly set the radius $R$ of the $S^3$
to one.  The dimensions of quantities such as energy
and temperature can be restored by affixing the appropriate
power of $R$.

We start by calculating $\Z(x)$ at tree
level, so the dilatation operator
reduces to $D = D_0$, which just counts the engineering
dimension of an operator.
To calculate \zx\ all we need to do is write down a complete basis
of operators and count them.
The most general gauge-invariant operator can be written as
a linear combination of operators with a definite number of traces,
and the most general $k$-trace operator can be expressed as a product
of $k$ single-trace operators (keeping in mind that separate traces
behave as identical particles and are subject to the
appropriate Bose or Fermi statistics).
Therefore, a complete basis for arbitrary gauge-invariant operators
follows naturally after we specify a complete basis of single-trace
operators.

\subsec{The $\N = 4$ alphabet}

A single-trace operator is a product of the fields
$(\phi^I, \lambda^a, F_{\mu \nu})$ of $\N = 4$ SYM theory and their
covariant derivatives.
Covariant derivatives must always be symmetrized with traces removed,
since antisymmetric derivatives can be replaced by field strengths
and traces of derivatives give terms which are zero by the equations
of motion.
Following Polyakov \PolyakovAF\ we refer to such objects
as the `letters' of the $\N = 4$ `alphabet'. We will use
$\A$ to denote the collection of letters.

We define $d(A)$ of a letter $A$ to be its engineering
dimension, so $d(\phi^I) = 1$, $d(\lambda^a) = {3 \over 2}$, $d(F_{\mu \nu})
= 2$, and each covariant derivative adds one to the dimension.
The enumeration of letters weighted by dimension gives
rise to the following elementary partition function for the
$\N = 4$ alphabet \refs{\SundborgUE,\AharonySX},
\eqn\zelem{{
z(x) = \sum_{A \in {\cal{A}}} x^{d(A)} =
{2 x (3 - \sqrt{x}) \over (1 - \sqrt{x})^3}.
}}
This function has a power series which converges for all
temperatures
$0 \le x < 1$ and can be understood as follows:
\eqn\zexpansion{
z(x) = \underbrace{6 x}_{6(\phi^I)} +
\underbrace{16 x^{3/2}}_{16(\lambda^a)}
+ \underbrace{30 x^2}_{24(D_\mu \phi^I) + 6(F_{\mu \nu})}
+ \underbrace{48 x^{5/2}}_{48(D_\mu \lambda^a)}
+ \underbrace{70 x^3}_{54(D_\mu D_\nu \phi^I) + 16(D_\mu F_{\nu\rho})}+
\cdots.
}
There are only 48 $D_\mu \lambda^a$ instead of 64 because
16 components are set to zero by the Dirac equation.  Similarly there
are only 54 components of $D_\mu D_\nu \phi^I$ because $D^2 \phi^I$
is zero.  Finally, 4 of the 24
components of $D_\mu F_{\nu \rho}$ are zero
by the equations of motion and another 4 are equal to zero by
the Bianchi identity.

In what follows we will frequently need to know various
partition functions with the fermion number
operator $(-1)^F$ inserted.
These factors can be easily dealt with by making use of the
fact that bosonic and fermionic operators respectively
have integer or half-integer dimensions at tree level.
To this end we introduce the quantity
\eqn\aaa{
\omega = e^{2 \pi i},
}
which is equal to $+1$ if it is raised to an integer power
and $-1$ if it is raised to a half-integer power.
To see $\omega$ in action consider the formula
\eqn\aaa{
z(\omega x) = \sum_{A \in \A} e^{2 \pi i d(A)} x^{d(A)}
 = {2 x (3 + \sqrt{x})
\over (1 + \sqrt{x})^3}
= 6 x - 16 x^{3/2} +  \cdots = \sum_{A \in \A}
(-1)^{F(A)} x^{d(A)}.
}

\subsec{Single-trace operators}

The next step is to string 
individual letters of the alphabet $\A$ together
to form single-trace operators.  The quantity we would like
to calculate is
the single-trace partition function
\eqn\aaa{
Z(x) = \sum_{{\rm single-trace}~O} x^{D_0(O)}.
}
(Note here the notational distinction between
the single-trace partition function $Z(x)$ and the full
partition function in \zx, denoted $\Z(x)$.)
The only constraint on $O = \Tr[ A_1 A_2 \cdots A_L]$ is
the overall cyclic invariance of the trace.
The enumeration of such single-trace operators is identical to
the combinatorial problem of counting the number of distinct necklaces
composed of a collection of different types of beads.
The solution to this problem, which we
review in the next section, may be expressed in terms
of $z(x)$ as
\eqn\singletracez{
Z(x)  
= - z(x) - \sum_{n=1}^\infty {\phi(n) \over n}
\ln[1 - z(\omega^{n+1} x^n)],
}
where $\phi(n)$ is Euler's totient function which counts the
number of integers less than $n$ which are relatively prime to $n$.
The first term $-z(x)$ in \singletracez\ is present simply to
subtract away traces of a single letter $\Tr[A]$, since these vanish
automatically in the $SU(N)$ theory.

Plugging \zelem\ into \singletracez\ gives an expansion which
goes as follows:
\eqn\aaa{\eqalign{
Z(x) &= 
\underbrace{21 x^2}_{21 (\Tr[\phi^I \phi^J])}
 + \underbrace{96 x^{5/2}}_{96(\Tr[\phi^I \lambda^a])}
\cr
&\qquad\qquad+
\underbrace{376 x^3}_{76(\Tr[\phi^I \phi^J \phi^K]) + 144(\Tr[\phi^I D_\mu
\phi^J]) + 36(\Tr[\phi^I F_{\mu\nu}]) + 120(\Tr[ \lambda^a \lambda^b])} +
\cdots.
}}
Two comments are in order.  The first is that the result
\singletracez\ is only valid in the $N = \infty$ limit of the
the $SU(N)$ gauge theory, since we allow arbitrarily high powers
of the individual letters.  If $N$ were finite, then trace identities
would allow a single trace of more than $N$ letters to be reexpressed
in terms of higher-trace operators,
indicating that the basis of operators we are using
would be overcomplete.

The second observation is that the power series expansion
of \singletracez\ converges
only for $0 \le x < x_{\rm H}$, where
\eqn\xhagedorn{
x_{\rm H} = 7 - 4 \sqrt{3} \approx 0.072.
}
The divergence of the partition function at this value, which corresponds
to the temperature
\eqn\aaa{
T_{\rm H}(\lambda=0) = - {1 \over \ln x_{\rm H}}=
 1/\ln(7 + 4 \sqrt{3}) \approx 0.380,
}
(measured in units of $R^{-1}$)
has been argued to be the gauge theory dual of the Hagedorn
transition in string theory \refs{\SundborgUE,\AharonySX}.
Hagedorn-like behavior in other free large $N$ systems
was also observed in \HalpernFZ.

\subsec{From single-trace to multi-trace operators}

Having determined the partition function $Z(x)$
for single-trace
operators, it is an easy combinatoric problem to calculate from
this the partition function for an arbitrary number of traces,
since the only constraint is that traces should be treated as
indistinguishable bosons or fermions.
The result is
\eqn\Zn{
\Z(x) =  \exp \left[ \sum_{n=1}^\infty
{ Z(\omega^{n+1} x^n) \over n} \right].
}
The partition function $Z_k(x)$ for $k$-trace operators can be extracted
by inserting $y^n$ into the sum and then reading off the coefficient
of $y^k$ in the expansion of the exponential.

{}From the
results reviewed above
we can see that
the complete partition function of $\N = 4$ $SU(N)$ gauge theory
at infinite $N$ and zero Yang-Mills coupling is given by
the formula
\eqn\zfree{
{\cal{Z}}(x) = 
\exp \left[ -
\sum_{n=1}^\infty
{ z(\omega^{n+1} x^n) \over n}
\right]
\prod_{n=1}^\infty
{1 \over 1 - z(\omega^{n+1} x^n)}.
}
The exponential term in \zfree\ is (one over) the partition function
for gauge group $U(1)$ and the infinite product is the partition function
for gauge group $U(N)$,
so \zfree\ expresses the expected fact
that
\eqn\aaa{
\Z_{SU(N)} = {\Z_{U(N)} \over \Z_{U(1)}}.
}
In fact, it is interesting
to note that although this analysis has been done at infinite
$N$, the result \zfree\ remains correct to all orders in $1/N$
(though certainly not at finite $N$).
This is true because
the tree-level dilatation operator $D_0$ obviously 
does not receive any $1/N$ corrections, and trace relations
are non-perturbative in $1/N$.

\subsec{Turning on the Yang-Mills coupling}

In \zfree\ we have reviewed the complete partition function
for $SU(N)$ $\N = 4$ SYM theory on $\R \times S^3$
at infinite $N$ and zero Yang-Mills coupling.
The goal of this paper is to calculate the first-order correction
to this result when we turn on the Yang-Mills coupling $g_{\rm YM}$.
The dilatation operator $D$ can be expanded in powers of
the 't Hooft parameter $\lambda = g_{\rm YM}^2 N$ as
\eqn\aaa{
D = D_0 + {\lambda \over 4 \pi^2} D_2 + \cdots,
}
so to first order in $\lambda$ the partition function \zx\ is given by
\eqn\aaa{
\Z(x,\lambda) = \Tr[x^{D_0 + {\lambda \over 4 \pi^2} D_2 + \cdots}]
= \Tr[x^{D_0}] + {\lambda \ln x \over 4 \pi^2} \Tr[x^{D_0} D_2]
+ \cdots.
}
Therefore we need to calculate the trace of the one-loop dilatation operator,
\eqn\tocalculate{
\Z^{(1)}(x) = \Tr[x^{D_0} D_2].
}
Our calculation will proceed by first calculating the one-loop
partition function in the single-trace sector and then assembling
together the result for multi-trace operators as in the previous
subsection.

\subsec{Operator mixing and $1/N$}

The tree-level dilatation operator $D_0$ is diagonal in the trace
basis, since the engineering dimension of a $k$-trace operator
is obviously just the sum of the engineering dimensions of the $k$
individual operators.
At one loop this is no longer true.  Instead
we have a formula of the form
\eqn\oneovern{
D_2 |k\rangle = \sum_{i=1}^k \Tr[O_1]
\cdots (D_2 \Tr[O_i]) \cdots \Tr[O_k] + {1 \over N} |k - 1\rangle
+ {1 \over N} |k+1\rangle
}
for a generic $k$-trace operator
$|k\rangle = \Tr[O_1] \cdots \Tr[O_k]$.
At first glance one might be tempted to disregard the ${\cal{O}}(1/N)$
terms since we are working at infinite $N$.  However it is well-known
that there are classes of $k$-trace
operators whose one-loop anomalous dimensions
receive non-zero contributions from mixing with $k\pm 1$-trace
operators \refs{\BeisertBB,\ConstableVQ}.
This occurs generically for BMN operators, which consist
of $L \sim \sqrt{N}$ letters.  For such operators the $1/N$ suppression
is overwhelmed by the growth of the number of non-planar
diagrams \ConstableHW.

Therefore we will not use $N = \infty$ as a justification to omit the
last two terms in \oneovern.  Fortunately we have an even
better excuse, which is simply that these terms are
non-diagonal in the trace
basis and therefore do not contribute to the quantity \tocalculate\ that
we are computing.  The situation becomes more complicated starting
at two loops, where the correction to the partition function involves
$\Tr[x^{D_0} D_4]$ and $\Tr[x^{D_0} D_2^2]$, both of which have
some diagonal terms of order $1/N^2$ which can't necessarily be dropped.
(In some sense this smells like an `order of limits' problem,
like the one discussed in \KlebanovMP).
Of course, if one focuses on calculating the planar partition
function (as opposed to the large $N$ partition function),
then none of these issues arise and one can ignore the $1/N$ terms
in \oneovern\ from the beginning.
Subtleties in the $1/N$ expansion have been noted in \BeisertTQ.

Note that we have implicitly been using a scalar product on the space of
operators which is diagonal in the trace basis,
\eqn\aaa{
\langle k | l \rangle \propto \delta_{kl}.
}
The  ${\cal{N}} = 4$ SYM theory provides, via the
state-operator correspondence, a natural inner
product $S_{kl} = \langle k|l\rangle$
on the space of operators which is diagonal in the
trace basis at infinite $N$ but receives non-diagonal,
trace-mixing corrections
beginning at ${\cal{O}}(1/N)$.
This operator mixing does not concern us here since the trace
\eqn\aaa{
\Tr[x^{D_0} D_2] = \sum_{k,l} \langle k|x^{D_0} D_2|l\rangle
S_{kl}^{-1}
}
is completely independent of the scalar product $S$.
Therefore we are free to choose the most convenient inner product
$S_{kl} = \delta_{kl}$.
A non-trivial inner product $S$ appeared in studies
of $1/N$ corrections to the BMN correspondence, such as
\refs{\GrossMH,\PearsonZS} where matrix elements of the Hamiltonian
were compared between gauge theory and string theory.
The utility of ignoring the gauge theory inner product for
the purpose of calculating basis-independent quantities was emphasized
in \BeisertFF\ in the context of calculating eigenvalues of $D_2$ in
the BMN sector.

\newsec{P\'olya Necklaces}

In this section we develop the machinery which will reduce the calculation
of the one-loop partition function
\eqn\toguess{
Z^{(1)}(x) = \Tr[x^{D_0} D_2]
}
in the single-trace sector to the calculation of some elementary
traces involving $\widehat{O} = D_2$
acting on two neighboring letters.
Roughly speaking, we consider here the problem of `tracing out' all but
two letters of any gauge-invariant operator, and express the result
in terms of traces of $\widehat{O}$ over the remaining two letters.
This is accomplished by translating the calculation into the
language of spin chains, and then using a generalization of
P\'olya's counting theory.

\subsec{Free necklaces}

A necklace $N$ of length $L$ is a collection of $L$ objects
$(A_1\cdots A_L)$ chosen from some fixed set $A \in \A$ of `beads',
such that two necklaces are identified if they differ from each
other by a cyclic rotation.
It is useful to introduce a counting function $d$ on the beads,
and define $d$ to act additively on the beads of a necklace,
\eqn\aaa{
d(A_1 \cdots A_L) = \sum_{i=1}^L d(A_i).
}
The analysis of this section will be general, but
of course for the desired application to $\N = 4$ SYM theory
we remember in the back of our mind that we will take $\A$ to be
the $\N = 4$ alphabet, and $d(A)$ will be the engineering
dimension of the letter $A$.

A central result of P\'olya's counting theory is that the necklace
partition function
\eqn\zneck{
Z(x) = \sum_{N} x^{d(N)},
}
where we sum over necklaces $N$ of arbitrary length $L \ge 1$,
is given by
\eqn\polya{
Z(x) = - \sum_{n=1}^\infty {\phi(n) \over n} \ln[1 - z(x^n)],
}
where $z(x)$ is the generating function for the beads,
\eqn\zbeads{
z(x) = \sum_{A \in \A} x^{d(A)}.
}
The formula \polya\ is valid when the beads are all bosons.
The generalization to include fermions is straightforward and will
be presented below.

\subsec{Necklaces with a pendant}

Instead of reviewing the elementary derivation of the result \polya,
we will consider a useful generalization from which we will recover
\polya\ as a special case.
To describe the generalization that we are interested in,
it is useful to think of a necklace of length $L$ as a spin
chain of length $L$, where on each site of the spin chain the
spin vector takes values in the set $\A$.
Of course we have
the constraint that only cyclically invariant spin configurations
correspond to necklaces.  Therefore, we can recast the calculation
of \zneck\ into spin chain language by writing the partition function as
\eqn\znecktwo{
Z(x) = \sum_N x^{d(N)} = \sum_{L=1}^\infty \Tr_L[ {\cal{P}} x^{d} ],
}
where $\Tr_L$ denotes the trace over spin chains of length $L$
and ${\cal{P}}$ denotes the projection operator onto the subspace
of cyclically
invariant
spin configurations.  The projector can be written explicitly as
\eqn\pdef{
{\cal{P}} = {1 \over L} (1 + T + T^2 + \cdots + T^{L - 1}),
}
where $T$ is the translation operator which sends site $i$ on
the chain to site $i+1$ and satisfies $T^L = 1$.

The generalization of $\Tr_L[{\cal{P}} x^d]$ that we would like to
consider is to the case
\eqn\generalization{
\Tr_L[ {\cal{P}} x^d \widehat{O}],
}
where $\widehat{O}$ is any homogeneous operator
which commutes with $d$ and 
which acts only
on two neighboring
sites at a time, so that it may be decomposed into the form
\eqn\odef{
\widehat{O} = \sum_{i=1}^L \widehat{{O}}_{i,i+1}, \qquad
\widehat{O}_{i,i+1} = 1_1 \otimes \cdots \otimes  1_{i-1}
\otimes {O} \otimes 1_{i+2} \otimes \cdots \otimes 1_L.
}
Spin chain Hamiltonians are of course prime examples of such
operators,
and eventually we will apply the present machinery to the case
$\widehat{O} = D_2$,
but our analysis will continue to be as general as possible.
Since the operator $O$ only connects two neighboring sites at a time,
but can slide all the way around the necklace, we can think of $O$
as a `pendant' hanging from two adjacent beads on the necklace.

Given any such operator $O$, an obvious quantity of interest
is the expectation value
\eqn\oexpectation{
\langle O(x) \rangle = \Tr_{\A \times \A}
[x^{d} O]
=  \sum_{A_1,A_2 \in \A}
x^{d(A_1) + d(A_2)} \langle A_1 A_2 | O | A_1 A_2\rangle.
}
We will see that knowing $\langle O(x)\rangle$
is almost, but not quite enough information to allow for the
calculation of 
\generalization.
The other quantity that we will need to know is the `permuted trace'
\eqn\poexpectation{
\langle PO(w,y)\rangle =
\Tr_{\A \times \A} [ P w^{d_1} y^{d_2} O ] =\sum_{A_1,A_2 \in \A}
w^{d(A_1)} y^{d(A_2)}
 \langle A_1 A_2 | O | A_2 A_1\rangle,
}
where $P$ is the permutation operator on $\A \times \A$ and
the two sites are counted with different variables $w$ and $y$.
Let us now see how to reduce the calculation of \generalization\ to
the calculation of these two quantities.

Since ${\cal{P}}$ projects onto cyclically invariant states anyway,
the sum over $i$ in \odef\ is actually redundant.
We can affix the pendant to sites $(i,i+1) = (1,2)$, and the sum in
\odef\ just gives
a factor of $L$ which cancels the $1/L$ in \pdef\ giving
\eqn\aaa{
\Tr_L[{\cal{P}} x^d \widehat{O}] =
\sum_{k=0}^{L-1} \Tr_L[ T^k x^d \widehat{O}_{12}].
}
This trace may be expressed as
\eqn\express{
\eqalign{
\Tr_L[ {\cal{P}} x^d \widehat{O} ]
&= \sum_{k=0}^{L-1} \sum_{A_1,\ldots,A_L}
\langle A_1 \cdots A_L| T^k x^d \widehat{O}_{12}
|A_1 \cdots A_L\rangle\cr
&=
\sum_{k=0}^{L-1} \sum_{A_1,\ldots,A_L}
x^{d(A_1) + \cdots + d(A_L)} \langle A_1 A_2| O |A_{1+k} A_{2+k}\rangle
\prod_{i=3}^L \delta_{A_i,A_{i+k}}.
}}

Upon contemplating the formula \express, it is clear that after
we sum over $A_3,\ldots,A_L$, only two possible structures can
emerge,
\eqn\twotype{
(i) ~~ \langle A_1 A_2 | O |A_1 A_2\rangle, \qquad {\rm or}
\qquad (ii) ~~ \langle A_1 A_2 | O |A_2 A_1\rangle,
}
depending on whether $k$ and $L$ are such that
the Kronecker delta functions in \express\ end up connecting
site $1+k$ to site 1 or to site 2.
For example, $k=0$ clearly gives the former structure
while
$k=1$ clearly gives the latter.
But for general $k$ and $L$, what is the criterion which tells
us which of the two possibilities \twotype\ we get?

Let us start with site $1 + k$ and follow it around the necklace
using $n$ of the Kronecker delta functions,
\eqn\utwo{
1 + k \to 1 + 2 k \to 1 + 3 k \to \cdots \to 1 + (n + 1) k.
}
Since there are only $L - 2$ delta functions in total, we have
$0 \le n \le L - 2$.
Now let $m = (L,k)$ be the greatest common divisor of $L$ and $k$.
If $m > 1$, then by step $n = L/m - 1$ at the latest,
site $1 + k$ 
will have been connected
to site
\eqn\aaa{
1 + k \to 1 + (n+1)k = 1 + (L/m - 1 + 1) k = 1 + (k/m) L = 1~{\rm mod}~L.
}
On the other hand, if $k$ and $L$ are relatively prime ($m=1$),
then there is no $n < L$ such that $1 + (n + 1) k = 1~{\rm mod}~L$, so
$1 + k$ cannot get connected to site 1 and hence must be connected to
site 2.
We conclude that we get the trace structure of type $(ii)$ if and
only if $k$ and $L$ are relatively prime, and we get structure $(i)$
otherwise.
Let us therefore study these cases separately.

\subsec{Case $(i)$: $k$ and $L$ have a common factor}

As a warm-up exercise let us consider the case $L = 15$, $k = 6$,
so that $m = (15,6) = 3$.
Then as we sum over sites $3,4,\cdots,L$, the delta functions in
\express\ sew together the beads of the necklace as follows:
\eqn\cycles{
\eqalign{
&7 \to 13 \to 4 \to 10 \to 1,\cr
&8 \to 14 \to 5 \to 11 \to 2,\cr
&9 \to 15 \to 6 \to 12 \to 3 \to 9.
}
}
Each line in this formula can be though of as a `strand' of the
necklace.  So the necklace with $L=15$ and $k=6$ is composed of
$m = (15,6) = 3$ strands, and the Kronecker delta functions
in \express\ force all of the beads on any given strand to
be the same.

The strands starting with $1 + k$ and $2 + k$ end up respectively
at sites 1 and 2, confirming our earlier analysis.
The third line in \cycles\ denotes the following contribution to
\express:
\eqn\aaa{
\eqalign{
&\sum_{A_3, A_9, A_{15}, A_6, A_{12}}
x^{d(A_3) + d(A_9) + d(A_{15}) + d(A_6) + d(A_{12})}
\delta_{A_3,A_9} \delta_{A_9,A_{15}}
\delta_{A_{15},A_6} \delta_{A_6,A_{12}} \delta_{A_{12},A_3}
\cr
&\qquad= \sum_{A_3} x^{5 d(A_3)} = z(x^5).
}}
The first two strands in \cycles\ involve the sites 1 and 2 where
the pendant is attached, and it is not hard to see that they end
up contributing a factor of
\eqn\aaa{
\sum_{A_1,A_2} x^{5 d(A_1) + 5 d(A_2)} \langle
A_1 A_2| O |A_1 A_2 \rangle = \langle O(x^5) \rangle,
}
using the definition \oexpectation.
Combining these results, we find that the total contribution
to the sum \express\ for the case $L=15$ and $k=6$ is
\eqn\aaa{
z(x^5) \, \langle O(x^5)\rangle.
}

The generalization of this analysis is straightforward.
A necklace
with general $L$ and $k$ will have $m = (L,k)$ different strands, each with
$L/m$ beads.  Two of those strands (like the first two
in \cycles) will involve the sites 1 and 2 and give rise to
\eqn\aaaaaa{
\langle O(x^{L/m}) \rangle.
}
The remaining $m - 2$ strands (note that we are assuming here
that $m \ge 2$), like the last line in \cycles, give a factor of
\eqn\aaabbb{
(z(x^{L/m}))^{m-2}.
}
Combining \aaaaaa\ and \aaabbb, we conclude that the total
contribution to \express\ from all $k$ such that $m = (k,L) > 1$
is
\eqn\contribone{
\sum_{k = 0 \atop m = (k,L) > 1}^{L-1}
\left[ z(x^{L/m})\right]^{m-2} \langle O(x^{L/m}) \rangle.
}

\subsec{Case $(ii)$: $k$ and $L$ are relatively prime}

As a warm-up exercise let us consider the case $L = 14$, $k = 5$.
The Kronecker
delta functions in \express\ now sew together the beads
\eqn\cyclestwo{\eqalign{
&6 \to 11 \to 2,\cr
&7 \to 12 \to 3 \to 8 \to 13 \to 4 \to 9 \to 14 \to 5 \to 10 \to 1,
}}
confirming our general analysis that site $1+k$ gets connected
to 2, and site $2 + k$ gets connected to 1,
giving trace structure $(ii)$.
The first line indicates a contribution of
\eqn\aaa{
\sum_{A_6,A_{11}} \delta_{A_6,A_{11}} \delta_{A_{11},A_2}
x^{d(A_6) + d(A_{11}) + d(A_2)} = x^{3 d(A_2)},
}
while the second line similarly denotes a contribution of
$x^{11 d(A_1)}$.  Putting everything together, we find
that for $L=14$ and $k=5$, \express\ reduces to
\eqn\aaa{
\sum_{A_1,A_2} x^{11 d(A_1) + 3 d(A_2)} \langle A_1 A_2|O|A_2 A_1\rangle
= \langle P O(x^{11}, x^3)\rangle,
}
using the definition \poexpectation.

The generalization to arbitrary $k$ and $L$ is straightforward.
After $n - 1$ steps in the first line of \cyclestwo, site $1 + k$
will connect to site $1 + n k$.  Therefore, the length of the first
strand is the smallest non-negative $n$ such that $1 + n k =2~{\rm
mod}~L$, or equivalently $n k = 1~{\rm mod}~L$.
The total contribution to
\express\ from all $k$ which are relatively prime to $L$ is
therefore
\eqn\becomes{
\sum_{k = 0\atop (k,L) = 1}^{L - 1}
\langle P O(x^{L-n(L,k)}, x^{n(L,k)}) \rangle, \qquad
n(L,k) = {\rm min}\{n \ge 0 : n k = 1~{\rm mod}~L\}.
}
Solving $nk = 1~{\rm mod}~L$ for fixed $k$ and $L$ is equivalent
to finding $n$, $m$ such that $n k - L m = 1$.
Given that $k$ is relatively prime to $L$, it is clear that
a solution exists only if $n$ and $L$ are also relatively prime.
Moreover, it is clear that $n \le L$ (otherwise subtracting $L$
from $n$ would give a smaller solution).
Finally, the set of $\{k : (k,L) = 1\}$ is in one-to-one correspondence
with the set of $n(L,k)$, simply because the condition $nk=1~{\rm mod}~L$
is symmetric in $n$ and $k$.
Therefore, although $n(L,k)$ is not generically equal to $k$,
the sum in \becomes\ is equivalent to\foot{We are grateful to
Jan Plefka for pointing out a flaw in our original proof of this
formula.}
\eqn\contribtwo{
\sum_{k = 0 \atop (k,L) = 1}^{L-1}
\langle P O(x^{L-k},x^{k})\rangle.
}

\subsec{Summary and main result}

We now combine the contributions \contribone\ and \contribtwo,
writing the result as
\eqn\answer{
\eqalign{
\Tr_L[{\cal{P}}x^d \widehat{O}]&= \sum_{k=0}^{L-1}
\left[ z(x^{L/(k,L)}) \right]^{(k,L) - 2}
\langle O(x^{L/(k,L)}) \rangle
\cr
&\qquad\qquad\qquad\qquad + \sum_{k=0 \atop (k,L) = 1}^{L-1}
\left[
\langle PO(x^{L-k}, x^{k})\rangle - z(x^L)^{-1}
\langle O(x^L)\rangle\right],
}}
where in the first term we omitted the constraint $m > 1$ from
\contribone\ at the expense of subtracting off the extra terms
on the second line.
Now we can trade the sum over $k$ in the first term of \answer\ for
a sum over divisors $a$ of $L$,
to write
\eqn\resone{
\Tr_L[{\cal{P}} x^d \widehat{O}]
= \sum_{a | L}
\phi(a) [ z(x^a) ]^{L/a - 2} \langle O(x^a) \rangle
+ \sum_{k = 0\atop (k,L) = 1}^{L-1}
\left[\langle P O(x^{L-k},x^{k})\rangle - z(x^L)^{-1}
\langle O(x^L)\rangle\right].
}
At this step let us pause for a moment to explain how to recover
the P\'olya formula \polya\ as promised.
To this end we need to consider the special case where the
operator $O$ is proportional to the identity matrix, and
specifically $O = 1/L$.  This is the correct normalization
which gives rise to $\widehat{O} = 1$ when plugged into the sum over
sites in \odef.
For $O = 1/L$ we easily find
\eqn\aaa{
\langle O(x) \rangle = {1 \over L} z(x)^2, \qquad
\langle P O(w,y) \rangle = {1 \over L} z(w y).
}
The second term in \resone\ therefore drops out, leaving just
\eqn\aaa{
\Tr_L[ {\cal{P}} x^d] = {1 \over L} \sum_{a | L}
\phi(a) [z(x^a)]^{L/a}.
}
The sum over $L$ is performed in the usual way, and we obtain
\eqn\aaa{
\sum_{L=1}^\infty \Tr_L [ {\cal{P}} x^d ]
= - \sum_{n=1}^\infty {\phi(n) \over n}
\ln[1 - z(x^n)],
}
which is the desired result \polya.

Having confirmed that the formula
\resone\ reduces to the known answer for the special case
$\widehat{O} = 1$, let us now consider operators
$O$ which do not depend explicitly on $L$.  In particular this
implies, through \odef, that the eigenvalues of $\widehat{O}$ scale
linearly with $L$.  Then we can perform the sum over $L > 1$ to arrive 
at
\eqn\prelimpolyaresult{
\eqalign{
\sum_{L=2}^\infty \Tr_L[{\cal{P}} x^d \widehat{O} ]
&= - {\langle O(x) \rangle \over z(x)}
+ \sum_{n=1}^\infty \phi(n) {\langle O(x^n)\rangle\over z(x^n)}
{1 \over 1 - z(x^n)}
\cr
&\qquad\qquad\qquad\qquad
+ \sum_{L=2}^\infty \sum_{k = 0\atop (k,L) = 1}^{L-1}
\left[
\langle PO(x^{L-k},x^{k})\rangle - { \langle O(x^L)\rangle \over
z(x^L)}\right].
}}
The first term is present to subtract off the part of the second term
which would correspond to $L=1$, which we omit since it can't support
a pendant (and moreover
is irrelevant in the $SU(N)$ gauge theory).
A final step is to simplify \prelimpolyaresult\ by changing the summation
variable from $n$ to $L$ and combining everything into
the main result
\eqn\mainpolyaresult{
{
\sum_{L=2}^\infty \Tr_L[{\cal{P}}x^d \widehat{O}]
= \sum_{L=1}^\infty
\sum_{k:(k,L) = 1}
\left[ {\langle O(\omega^{L+1} x^L)\rangle \over 1 - z(\omega^{L+1} x^L)}
+ \delta_{L \ne 1}
\langle PO(\omega^{L-k+1} x^{L-k},\omega^{k+1} x^{k}\rangle\right].
}
}
It is a straightforward exercise to generalize the analysis
of the previous subsections to allow for fermionic beads,
and we have included here the appropriate factors of $\omega$
which keep track of the minus signs appearing when such beads are permuted.
(The permutation operator $P$ is understood to be graded,
i.e.~$P|A_1A_2\rangle = (-1)^{F_1 F_2}|A_2 A_1\rangle$.)

The first term in \mainpolyaresult\ for $L=1$ is
precisely what one would obtain by making the
crude estimate that the only effect of the projection ${\cal{P}}$
onto cyclically invariant states is to insert a factor of $1/L$.
The rest of
\mainpolyaresult\ 
is the detailed correction to this approximation.

In all the cases relevant to $\N = 4$ SYM theory that we
study below, we will see that the second
term
in \mainpolyaresult\ 
is a very small correction in the sense that its contribution
to the coefficient of $x^n$ is negligible for large $n$.
In particular, we will find that $\langle O(x)\rangle$
and $\langle PO(w,y)\rangle$ converge for all temperatures
so that the large temperature behavior of \mainpolyaresult\ is dominated
by the pole $1/(1 - z(x))$
in the first term.

\newsec{Examples}

We can gain some insight into the formula
\mainpolyaresult\ by applying it to some subsectors of the
gauge theory.
The implication of the results presented in
this section for the thermodynamics
of $\N = 4$ SYM theory is unclear since there is no sense in which
the sectors decouple from each other at finite temperature (we
do not consider here the addition of chemical potentials for
various charges).
However, we believe this section is a useful warm-up exercise for
the more complicated analysis which follows.  Moreover,
the results given here for the traces of the $SU(2)$, $SO(6)$
and $SL(2)$ spin chain Hamiltonians may be of interest from the
point of view of integrability in those sectors.
An additional subsector which is of independent interest is the
$SU(2|4)$ subsector, which at one loop is isomorphic to the
't Hooft large $N$ limit of the plane-wave matrix model \BerensteinJQ.
This subsector is considered in \MMP.

\subsec{The $SU(2)$ subsector}

This subsector consists of all operators of
the form
\eqn\aaa{
\Tr[X X Z Z Z X Z Z  \cdots X X Z],
}
where $X$ and $Z$ are two holomorphic scalar fields.
The `alphabet' for this sector is $\A = \{ X, Z \}$,
the dimension formula is $d(A) = 1$, and
the elementary partition function  is $z(x) = 2 x$.
In the natural basis $\{ XX, XZ, ZX, ZZ \}$
for $\A \times \A$, the matrix elements of the
permutation operator $P$ and the one-loop
Hamiltonian $D_2$ are simply
\eqn\aaa{
P = \pmatrix{1 & 0 & 0 & 0\cr
0 & 0 & 1 & 0 \cr
0 & 1 & 0 & 0\cr
0 & 0 & 0 & 1},
\qquad
D_2 = {1 \over 2} (1 - P).
}
We immediately find
\eqn\aaa{
\langle  D_2(x) \rangle = x^2, \qquad
\langle P D_2(w,y) \rangle = -  w y.
}
Plugging these into the main formula \mainpolyaresult\ gives,
after some simplification, the following formula for the
trace of the Hamiltonian in the $SU(2)$ sector:
\eqn\sutwo{
\Tr[x^{D_0} D_2] =  x
-\sum_{n=1}^\infty \phi(n) x^n \left({1 - 3 x^n \over 1 - 2 x^n}
\right).
}
As a check,
we used a computer to calculate the trace of the $SU(2)$ spin chain
Hamiltonian for all chains of length $L \le 26$ and successfully
matched the coefficients in the expansion of \sutwo\ up to order
$x^{26}$.

\subsec{The $SO(6)$ subsector}

This subsector consists of all operators built only out of scalar fields
with no derivatives,
\eqn\aaa{
\Tr[\phi^{I_1} \cdots \phi^{I_L}], \qquad I_i=\{ 1,\ldots,6 \}.
}
The alphabet is $\A = \{ \phi^1,\ldots,\phi^6\}$, the dimension
formula is $d(A) = 1$, and the elementary partition function is
$z(x) = 6 x$.
In the natural basis $|I_1 I_2 \rangle = \phi^{I_1} \phi^{I_2}$
for $\A \times \A$,
the matrix elements of $D_2$ and $P$ are \MinahanVE
\eqn\aaa{\eqalign{
\langle I_1 I_2 | D_2 | J_1 J_2 \rangle
&= {1 \over 4}
\delta_{I_1 I_2} \delta_{J_1 J_2}
+ \ha \delta_{I_1 J_1} \delta_{I_2 J_2} - \ha \delta_{I_1 J_2}
\delta_{I_2 J_1},\cr
\langle I_1 I_2 | P | J_1 J_2 \rangle
&= \delta_{I_1 J_2} \delta_{I_2 J_1}.
}}
A simple calculation yields
\eqn\aaa{
\langle D_2(x) \rangle = {33 \over 2} x^2, \qquad \langle P D_2(w,y)
\rangle = -{27 \over 2} w y,
}
which leads to
the result
\eqn\sosix{
\Tr[x^{D_0} D_2] =
{27 \over 2} x
- {3 \over 2} \sum_{n=1}^\infty \phi(n) x^n \left( {9 - 65 x^n \over
1 - 6 x^n} \right)
}
for the trace of the $SO(6)$ Hamiltonian.
As a check,
we used a computer to calculate the trace of the $SO(6)$ spin chain
Hamiltonian for all chains of length $L \le 11$ and successfully
matched the coefficients in the expansion of \sosix\ up to order
$x^{11}$.

\subsec{The $SL(2)$ subsector}

This subsector consists of all operators of the form
\eqn\aaa{
\Tr[ D^{i_1} Z \cdots D^{i_L} Z], \qquad i_n \in \{0,1,\ldots\},
}
where $Z$ is a single holomorphic scalar field and $D$ is a single
holomorphic covariant derivative.
The alphabet is $\A = \{ Z, D Z, D^2 Z, \cdots \}$, the dimension
formula is $d(D^i Z) = i + 1$,
and the elementary partition function is
\eqn\aaa{
z(x) = \sum_{A \in \A} x^{d(A)} = \sum_{i=0}^\infty x^{i+1} = {x \over 1-x}.
}
In the basis $|i_1 i_2\rangle = D^{i_1}\!Z\,D^{i_2}\! Z$
for $\A \times \A$ the matrix elements of $D_2$ are \BeisertJJ
\eqn\aaa{
\langle i_1 i_2 | D_2 | j_1 j_2 \rangle
= {1 \over 2}
\delta_{i_1 + i_2,j_1+j_2} \left[
\delta_{i_1,j_1} (h(j_1) + h(j_2)) -
{ \delta_{i_1 \ne j_1} \over | i_1 - j_1|}\right],
}
where $h(j)$ are the harmonic numbers
\eqn\aaa{
h(j) = \sum_{n=1}^j {1 \over n}, \qquad h(0) = 0.
}
The matrix elements of $P$ are obviously
\eqn\aaa{
\langle i_1 i_2 | P | j_1 j_2 \rangle
= \delta_{i_1 j_2} \delta_{i_2 j_1}.
}
A simple calculation yields
\eqn\aaa{
\langle D_2(x) \rangle
= - {x^2 \over (1 - x)^2} \ln(1-x)
}
and
\eqn\aaa{
\langle P D_2(w,y) \rangle
= {1 \over 2} {w y \over 1 - w y} \ln\left[{(1-w)(1-y) \over
(1 - wy)^2}\right].
}
After some simplification, we find
for the trace of the $SL(2)$ Hamiltonian the result
\eqn\sltwo{
\Tr[x^{D_0} D_2] = 
- \sum_{n=1}^\infty \phi(n) {  x^n \over 1 - 2 x^n}
\ln(1 - x^n)
+  \sum_{L=1}^\infty {x^L \over 1 - x^L}
\sum_{k = 1 \atop (k,L) = 1}^L
\ln (1 - x^k).
}
As a check,
we used a computer to calculate the trace of the $SL(2)$ spin chain
Hamiltonian for all chains
with total dimension $D_0 \le 19$ and successfully
matched the coefficients in the expansion of \sltwo\ up to order
$x^{19}$.

\newsec{Traces of the $PSL(4|4)$ Spin Chain Hamiltonian}

We now turn to
our next step, which is to apply the result \mainpolyaresult\ to
the calculation of the one-loop correction to 
the partition function of $\N = 4$ SYM theory on $\R \times S^3$
in the single-trace sector:
\eqn\aaa{
Z^{(1)}(x) = \Tr[x^{D_0} D_2].
}
To this end, we compute in this section the quantities
\eqn\aaa{
\langle D_2(x) \rangle
= \Tr_{\A \times \A} [x^{D_0} D_2],\qquad
\langle P D_2(w,y) \rangle
= \Tr_{\A \times \A} [ P w^{D_{0(1)}} y^{D_{0(2)}} D_2]
}
needed to invoke \mainpolyaresult\ for the full $PSL(4|4)$ spin
chain Hamiltonian $D_2$.

The calculation of $\langle D_2(x) \rangle$
is greatly facilitated by making use of the $PSL(4|4)$ symmetry
of $\N=4$ SYM theory.
Since the dilatation operator $D_2$ commutes with this symmetry, the action
of $D_2$ on an arbitrary state can be decomposed into its
action on irreducible representations of $PSL(4|4)$.
We therefore
begin with a discussion of the relevant representations.
Unfortunately, the operator
$w^{D_{0(1)}} y^{D_{0(2)}}$ does
not commute with the two-letter $PSL(4|4)$ Casimir, so
the calculation of $\langle P D_2(w,y) \rangle$ will prove
more difficult.

\subsec{Digraphs in the $\N = 4$ language}

The elementary fields and their covariant derivatives which
make up the alphabet $\A$ of $\N = 4$ SYM theory
constitute the so-called `singleton' representation of
$PSL(4|4)$.
The superconformal
primary state is the scalar field $\phi^I$, with
quantum numbers
\eqn\aaa{
[0,1,0]_{(0,0)}
}
under $SL(4)$ flavor rotations and the $SL(2) \times SL(2)$
Lorentz algebra.
The singleton representation is frequently denoted $\V_{\rm F}$,
although we shall continue to refer to it as $\A$ for consistency.

Since the one-loop dilatation operator $D_2$
only acts on two letters at
a time and commutes
with the two-letter Casimir $J^2_{(12)}$
of $PSL(4|4)$, it is sufficient to consider the
decomposition of the
product of
two copies of the singleton representation into
irreducible representations of $PSL(4|4)$.
The decomposition is
\eqn\decomposition{
{\cal{A}} \times {\cal{A}} = \sum_{j=0}^\infty \V_j,
}
where $\V_j$
is the module whose superconformal primary is
an eigenstate of the $PSL(4|4)$ Casimir with eigenvalue
$j(j+1)$ and quantum numbers
\eqn\aaa{\eqalign{
[0,2,0]_{(0,0)}, \qquad [1,0,1]_{(0,0)}, \qquad
{\rm and}\qquad
 [0, 0, 0]_{({j \over 2} - 1, {j \over 2} - 1)}~{\rm for}~j \ge 2.
}}
In the notation of \DolanZH, we have
\eqn\aaa{
\eqalign{
\A &= {\cal{B}}^{{1 \over 2}, {1 \over 2}}_{[0,1,0](0,0)},
\cr
\V_0& = {\cal{B}}^{{1 \over 2}, {1 \over 2}}_{[0,2,0](0,0)},
\cr
\V_1 &= {\cal{B}}^{{1 \over 4}, {1 \over 4}}_{[1,0,1](0,0)},
\cr
{\rm and}~\V_j &= {\cal{C}}^{1,1}_{[0,0,0]({1 \over 2} j - 1,
{1 \over 2} j - 1)}~{\rm for}~j \ge 2.
}}
In linguistics, a group of two successive letters whose phonetic
value
is a single sound, such as {\it ng} in {\it Yang} or {\it th}
in {\it theory}, is called
a `digraph', so we can think of the $\V_j$ as the digraphs
of $\N = 4$ Yang-Mills theory.

It is straightforward to count the primary states in $\V_j$,
weighted in the usual way by $x^{D_0}$.
For $j = 0$ and $j \ge 2$ the results can be read off
respectively
from (6.13) and (5.45) of \DolanZH, or
from tables 7  and 8 of \BianchiWX.
We did not immediately find the primary content of $\V_1$ in the
literature, but the derivation thereof is straightforward and the result
is presented in appendix A.
The results for all $\V_j$ can be summarized in the expressions
\eqn\vjdef{\eqalign{
V_0(x) &= 4 x^2 (1 + \sqrt{x})^4 (5-x),\cr
V_j(x) &= x^j (1 + \sqrt{x})^7 (j - 1 + (j + 2) \sqrt{x})(j - 1
+ 5 \sqrt{x} - (j+2) x), \qquad j \ge 1.
}}
Setting $x = 1$ counts the total number of primaries
in $\V_j$, which is $2^8 (2 j + 1)$ for any $j \ge 0$.

Note that for $j=0,1$ some powers of $x$ in $V_j(x)$
have negative coefficients.
This may be thought of as a bookkeeping device (explained in \DolanZH)
which allows for easily keeping track of fields which are eliminated
by the requirement of imposing equations of motion or conservation
laws.  One consequence of this is that the full partition function
for the module $\V_j$
(including descendants) may be calculated naively,
with derivatives treated as if they acted freely,
without worrying about equations of motion or conservation laws.
The partition function for the module $\V_j$ is therefore
simply
\eqn\trvj{
\Tr_{\V_j}[x^{D_0}]
= \Tr_{\A \times \A}[P_j x^{D_0}]
= { V_j(x) \over (1-x)^4},
}
where we have defined $P_j$ to be the projection
operator $P_j : \A \times \A \to \V_j$.

The $P_j$ form a complete set of orthogonal projection operators,
so
\eqn\aaa{
\sum_{j=0}^\infty P_j = 1.
}
This implies the identity
\eqn\aaa{
\sum_{j=0}^\infty \Tr_{\A \times \A} [P_j x^{D_0}]
=
\Tr_{\A \times \A} [x^{D_0}] = z(x)^2,
}
which is indeed satisfied by \vjdef\ and \trvj.

\subsec{Simple trace $\langle D_2(x)\rangle$}

Here we calculate the expectation value \oexpectation\ for
the one-loop dilatation operator $D_2$.
The calculation only takes one line since we have all of
the machinery in place.
In \BeisertJJ\ it was shown that the eigenvalue of $D_2$ in the
module $\V_j$ is simply the harmonic number
\eqn\aaa{
h(j) = \sum_{n=1}^j {1 \over n}, \qquad h(0) \equiv 0,
}
and therefore that $D_2$ may be written as
\eqn\dtwo{
D_2 = \sum_{j=0}^\infty h(j) P_j.
}
{}From \trvj\ and \dtwo\ we immediately have
\eqn\aaa{
\langle D_2(x) \rangle =
\Tr_{\A \times \A} [ x^{D_0} D_2]
=  \sum_{j=0}^\infty h(j) {V_j(x) \over (1 - x)^4}.
}
Plugging in \vjdef\ and performing the sum over $j$ gives
\eqn\dtwotrace{
{
\langle D_2(x)\rangle
= {(1 + \sqrt{x})^2 \over (1 - \sqrt{x})^6}
\left[ - (1 - 4 \sqrt{x} + x)^2 \ln(1-x) - x (1 - 8 \sqrt{x} + 2 x)\right].
}
}

\subsec{Permuted trace $\langle P D_2(w,y)\rangle$}

As mentioned above, the calculation of
$\langle P D_2(w,y)\rangle$ is complicated by the fact
the two-letter $PSL(4|4)$ Casimir operator $J^2_{(12)}$
commutes only with the sum $D_{0(1)} + D_{0(2)}$, but not
with $D_{0(1)}$ and $D_{0(2)}$ separately.
In particular, it does not commute with
$w^{D_{0(1)}} y^{D_{0(2)}}$, so the beautiful decomposition
into the modules $\V_j$ that we employed in the previous subsection is
of no use here.
This apparent breaking of $PSL(4|4)$
is merely an artifact of our choice to simplify
the calculation of $\Tr_L[{\cal{P}} x^{D_0} D_2]$
by tracing out $L-2$ sites on the chain and expressing
\mainpolyaresult\ in terms of the two remaining sites.

A manifestly $PSL(4|4)$-invariant calculation of
$\Tr_L[{\cal{P}} x^{D_0} D_2]$ would proceed as follows.
First, one would need to know the collection ${\cal{C}}_L$
of $PSL(4|4)$ modules which appear in higher powers
of the singleton representation $\A$,
\eqn\bigde{
\Tr(\A \times \cdots \times \A) =\Tr(\A^L)= \sum_{{\cal{I}} \in {\cal{C}}_L}
\V_{\cal{I}}
}
(where $\Tr$ denotes the projection onto singlets of the cyclic
group $\ZZ_L$).
Then $PSL(4|4)$ invariance guarantees that in each
resulting module $\V_{\cal{I}}$ the dilatation
operator $D_2$ is proportional to the identity operator, with
some calculable eigenvalue $h_{\cal{I}}$.
The desired trace would then be given by
\eqn\aaa{
\Tr [x^{D_0} D_2]
= \sum_{L=2}^\infty
\sum_{{\cal{I}} \in {\cal{C}}_L} h_{\cal{I}} \Tr[P_{\cal{I}} x^{D_0}],
}
where $P_{\cal{I}}$ is the projection operator from $\Tr(\A^L)$ onto
$\V_{\cal{I}}$.

The decomposition of $\A \times \A \times \A$ appears
in \BeisertDI\ (see also \BeisertRY),
where it was used to determine the one-loop anomalous dimensions
of some operators consisting of three elementary fields.
However,
it seems quite challenging to implement
the procedure outlined in the previous paragraph for arbitrary $L$,
although of course it would be very interesting to do so.

Instead, we will proceed by using
the matrix elements of the
operator $D_2$, written down
in \BeisertJJ\ in a $GL(4|4)$ oscillator basis, and
then calculating
the desired trace `by hand'.
This is quite a lengthy calculation, so we begin by presenting
the result.  The interested reader can find more details below.
We find:
\eqn\pdtwo{{\eqalign{
\langle P D_2(w,y)\rangle
&=
{1 + \sqrt{w} \over (1-\sqrt{w})^3}
{1 + \sqrt{y} \over (1-\sqrt{y})^3}
{1 \over (1 + \sqrt{w y})^3 (\sqrt{w} +
\sqrt{y})^2}
\Bigg[
w y (1-w)(1-y) p_1(w,y) \cr
&\qquad\qquad\qquad +{1 \over 2} \ln\left[{(1-w)(1-y) \over (1-w y)^2}\right]
(1 - \sqrt{w y}) (\sqrt{w} + \sqrt{y})^2 p_2(w,y)\cr
&\qquad\qquad\qquad -
{1 \over 2} { (\sqrt{w}-\sqrt{y})\over
(\sqrt{w}+\sqrt{y})} \ln\left[{1-w \over 1-y}\right]
(1 + \sqrt{w y})^3 p_3(w,y) \Bigg],
}
}}
in terms of the three polynomials
\eqn\aaa{\eqalign{
p_1 &= 4 - 16{{w}^{1/2}}
+ 7w - 16{{y}^{1/2}} + 22{{w}^{1/2}}{{y}^{1/2}} -
   16w{{y}^{1/2}}\cr
&\qquad\qquad + w^{{3/2}}{{y}^{1/2}} + 7y -
   16{{w}^{1/2}}y + 6wy + {{w}^{1/2}}y^{{3/2}},
}}
\eqn\aaa{
\eqalign{
p_2&=
 1 - 4{{w}^{1/2}}
 + w - 4{{y}^{1/2}} + 20{{w}^{1/2}}{{y}^{1/2}} -
   20w{{y}^{1/2}} + 4w^{{3/2}}{{y}^{1/2}} + y -
   20{{w}^{1/2}}y\cr
&\qquad\qquad + 42wy - 20w^{{3/2}}y + w^2y +
   4{{w}^{1/2}}y^{{3/2}} - 20wy^{{3/2}} +
   20w^{{3/2}}y^{{3/2}}\cr
&\qquad\qquad - 4w^2y^{{3/2}} + wy^2 -
   4w^{{3/2}}y^2 + w^2y^2,
}}
and
\eqn\aaa{
\eqalign{
p_3 &=
 w - 4w^{{3/2}} + w^2 + 4{{w}^{1/2}}{{y}^{1/2}} -
   20w{{y}^{1/2}} + 20w^{{3/2}}{{y}^{1/2}} - 4w^2{{y}^{1/2}} +
   y\cr
&\qquad\qquad - 20{{w}^{1/2}}y + 42wy - 20w^{{3/2}}y + w^2y -
   4y^{{3/2}} + 20{{w}^{1/2}}y^{{3/2}}\cr
&\qquad\qquad -
   20wy^{{3/2}} + 4w^{{3/2}}y^{{3/2}} + y^2 -
   4{{w}^{1/2}}y^2 + wy^2.
}}
Note the interesting identity
\eqn\aaa{
p_3(w,y) = y^2 p_2(1/y,w) = w^2 p_2(y,1/w).
}

It is useful to subject this complicated result to a simple
check.  When we set $w=y=x$, then the calculation
of $\langle P D_2(x,x)\rangle$
can be done using the group theoretic techniques of the previous subsection.
In particular, we have
\eqn\aaa{\eqalign{
\langle P D_2(x,x) \rangle &=
\Tr_{\A \times \A} [ P x^{D_0} D_2] \cr
&= \sum_{j=0}^\infty \Tr_{\V_j} [P x^{D_0} D_2] \cr
&= \sum_{j=0}^\infty (-1)^j h(j) { V_j(x) \over (1 - x)^4},
}}
where we used the fact that the permutation operator $P$ acts
as $(-1)^j$ in $\V_j$  \refs{\DolanUH,\DolanPS}.
If we now substitute the expressions for $V_j(x)$ from \vjdef\ and
perform the sum over $j$, we obtain
\eqn\diagp{
\eqalign{
&\langle P D_2(x,x)\rangle= 
{(1 + \sqrt{x})^3 \over
(1 - \sqrt{x})^4} {1 \over (1 + x)^3}\Big[
x (1 - 7 x^{1/2} + x + x^{3/2} - 6 x^2 + 2 x^{5/2})\cr
&\qquad\qquad-
(1 - 7 x^{1/2} + 15 x - 25 x^{3/2} + 25 x^2 - 15 x^{3/2} +7
x^3 - x^{7/2}) \ln(1+x)
\Big].
}}
Encouragingly, this expression agrees precisely with that obtained
by setting $w=y=x$ in \pdtwo.

Now we begin in earnest the calculation of \pdtwo.  The first
step is to use the matrix elements of $D_2$ in a $GL(4|4)$
oscillator basis, as presented in \BeisertJJ, to write down
a sum which gives \pdtwo:
\eqn\rone{\eqalign{
\langle P D_2(w,y)\rangle &=
\sum_{s_1,s_2,p_1,p_2,k=0}^\infty
\sum_{F_1,F_2,j=0}^4
(-1)^j {4 \choose j}
{4-j \choose F_1 - j}
{4 - j \choose F_2 - j}
{p_1! p_2! \over
k! (k+1)!}
\cr
&\qquad\times
c(n_1 + n_2, n_1-k-j, n_2-k-j)
w^{1 + s_1/2 + p_1/2}
y^{1 + s_2/2 + p_2/2}
\cr
&\qquad\times
{\cal{F}}(-1-k,-k,-s_1,-s_2;2,1-k+p_1,1-k+p_2;1) \cr
&\qquad \times \prod_{i=1}^2
\delta(1 - \half s_i + \half
p_i - \half F_i)
(1+s_i)(1+p_i),
}}
where ${\cal{F}}$ is the regularized hypergeometric function,
\eqn\aaa{
n_i = s_i + p_i + F_i,
}
and the coefficients $c$ are
the matrix elements of $D_2$ given in \BeisertJJ:
\eqn\coefficients{
\eqalign{
c(n,n_{12},n_{21}) &= {1 \over 2} (-1)^{1 + n_{12} n_{21}}
{ \Gamma(\half(n_{12} + n_{21})) \Gamma(1 + \half(n-n_{12}-n_{21}))
\over \Gamma(1 + \half n)},\cr
c(n,0,0)&= \half h(\half n).
}}
The detailed derivation of \rone, which is not entirely straightforward,
is presented in appendix B.
The next several steps of the calculation will be shown schematically.
The skeptical reader is free to verify that the power series
expansion of \rone\ agrees with that of \pdtwo\ to any desired order.

After several manipulations similar to the ones in appendix B,
equation \rone\ can be cast into the form
\eqn\rtwo{
\eqalign{
\langle P D_2(w,y)\rangle
&= \sum_{n=0}^\infty (-1)^n (n+1)(n+3) w^{1+n/2} y^{-n/2}
\cr
&\qquad\times
\left[ { (1 + \sqrt{y})^2 \over
(1 - \sqrt{y})^2 } P_n(y)
+ (1 + (-1)^{n+1} y^{1 + n/2})^2
{\ln(1-y) \over y} \right],
}}
where $P_n(y)$ is a polynomial in $\sqrt{y}$ whose
highest term is ${\cal{O}}(y^{n+1})$.
We did not obtain an explicit formula for $P_n(y)$, though presumably
it could be reverse engineered from the final answer \pdtwo.
Instead, we use a trick by investigating the quantity
\eqn\qdef{
Q(w,y) = \langle P D_2(w,y) \rangle
\left[ { (1 - \sqrt{y}) \over (1 + \sqrt{y})
}
{(1 - \sqrt{w}) \over ( 1 + \sqrt{w})
}
\right]^2.
}
Now we use \rtwo\ and the power series expansion
\eqn\aaa{
\left(
{1 - \sqrt{w} \over 1 + \sqrt{w}
}
\right)^2
= \sum_{m=1}^\infty d_m w^{m/2},
\qquad d_m = \delta_{m,0} + 4 m (-1)^m
}
to write
\eqn\qone{
Q(w,y) = \sum_{m,n=0}^\infty
d_m
(-1)^n (n+1)(n+3) w^{1+n/2+m/2} \left(
y^{-n/2} P_n(y) + W_n(y)\right),
}
where
\eqn\aaa{
W_n(y) = y^{-n/2} \left[ {1 - \sqrt{y} \over 1 + \sqrt{y}}\right]^2
(1 + (-1)^{n+1} y^{1+n/2})^2 {\ln(1-y) \over y}.
}

The trick is now to break the power series expansion of $Q(w,y)$
into the upper diagonal terms (where the power of $y$ is greater
than the power of $w$), the diagonal terms, and the lower diagonal terms.
Since $Q(w,y)$ is a symmetric function, the lower diagonal terms
will be known once the upper diagonal terms are known.
Furthermore, the diagonal terms can be extracted from \diagp, so
all we have left to calculate are the upper diagonal terms.
But since $P_n$ is a polynomial whose highest-order term is
$y^{n+1}$, we see from
\qone\ that it never contributes to the upper diagonal.
Therefore, for purposes of computing the upper diagonal
we are free
to omit the $P_n(y)$ term in \qone, and make the replacement
\eqn\aaa{
W_n(y) \to \left. W_n(y)\right|_{y^k: k > 1 + n/2 + m/2},
}
where the notation means that we write out $W_n(y)$ as a series in $y$
and throw away all powers of $y$ less than or equal to $1 + n/2 + n/2$.
This finally gives a formula for $Q(w,y)$ which is amenable to a
calculation in Mathematica.
In this manner, we obtain after a tedious but straightforward calculation
the result \pdtwo.

\newsec{One-Loop $\N = 4$ SYM Partition Function}

\subsec{Single-trace}

The complete one-loop correction to the partition function of $\N=4$
SYM theory on $\R \times S^3$ in the single-trace sector is given
by substituting \dtwotrace\ and \pdtwo\ into \mainpolyaresult.
We will not rewrite the formulas because no significant simplification
seems to occur.  Instead, let us note that the result
has the expansion
\eqn\singletrace{
{\lambda \ln x \over 4 \pi^2} \Tr[x^{D_0} D_2]
= {\lambda \ln x \over 4 \pi^2}
\left[ 3 x^2 + 48 x^{5/2} + 384 x^3 + 2064 x^{7/2} + \cdots\right].
}
(Each coefficient receives contributions
from both kinds of traces,
$\langle D_2\rangle$ and $\langle P D_2\rangle$.)
The first term in \singletrace\ encodes the one-loop anomalous dimension
of the Konishi operator, $3/4 \pi^2$.
The second term is $3 \times 16$, coming
from 16 descendants of the Konishi operator.  The third
term is $384 =  6 \times 2 + 20 \times 3 + 104 \times 3$,
which come respectively from 
the 6 primary states $\Tr[\phi^I \phi^I \phi^J]$, which have anomalous
dimension $1/2 \pi^2$ according to table 3 in \BeisertJJ,
the 20 Konishi descendants of the form $\Tr[\phi^I[\phi^J,\phi^K]]$,
and finally 104 Konishi descendants
which are traces of two elementary fields.

\subsec{Multi-trace}

Now let us extend the result of the previous subsection to
the complete one-loop correction to the partition function,
including multi-trace operators.
As discussed in section 2.6, the diagonal matrix elements of the one-loop
dilatation operator act additively on $k$-trace operators.
Therefore, to go from the single-trace partition function to the multi-trace
partition function
we can still use the formula \Zn.  Substituting 
\eqn\aaa{
Z(x) = Z^{(0)}(x) + {\lambda \ln x \over 4 \pi^2} Z^{(1)}(x)
}
and expanding to first order in $\lambda$, we find that the first
order correction to the multi-trace partition function is
\eqn\toplug{
\Z^{(1)}(x) = {\lambda \ln x \over 4 \pi^2}
\Z^{(0)}(x) \sum_{n=1}^\infty
Z^{(1)}(\omega^{n+1} x^n),
}
(a factor of $1/n$ is canceled by $\ln x^n = n \ln x$)
where we recall that the tree level result $\Z^{(0)}(x)$
is written in \zfree.

Now let us plug the result from \mainpolyaresult\ into \toplug.
The term proportional to $\langle P D_2 \rangle$ gives the sum
\eqn\tone{
\sum_{n=1}^\infty \sum_{L=2}^\infty \sum_{k:(k,L)=1}
\langle PD_2(
\omega^{n(L-k)+1} x^{n(L-k)},
\omega^{nk+1} x^{nk}
)\rangle
= \sum_{a,b=1}^\infty \langle P D_2(\omega^{b+1} x^b,
\omega^{a+1} x^a)\rangle.
}
To see why this equation is true, pick any positive $a$ and $b$, and 
look at the left-hand side to see how many times (if any)
the term $\langle P D_2(\omega^{a+1} x^a, \omega^{b+1} x^b)\rangle$
appears. This is equivalent to asking how many solutions, for given $a$
and $b$, there are to the equations
\eqn\eeeqn{
n k = a, \qquad n L = a + b
}
for $L \ge 2$ and $k: (k,L) = 1$.
The answer is that there is always precisely one solution:
$n = (a,b)$, $L = (a + b)/n$ and $k = a/n$.
Certainly if $n$ were not the greatest common divisor of $a$ and $b$
but some smaller common divisor, then \eeeqn\ would still give
solutions for $k$ and $L$, but these would not satisfy the
constraint $(k,L) = 1$.

Next we plug the $\langle D_2(x)\rangle$ terms from \mainpolyaresult\ into
\toplug,
which gives the sum
\eqn\ttwo{
\sum_{n=1}^\infty \sum_{L=1}^\infty \phi(L) { \langle D_2(
\omega^{nL+1} x^{n L})\rangle
\over 1 - z(\omega^{nL+1} x^{n L})}
= \sum_{k=1}^\infty \left[\sum_{L | k} \phi(L) \right]
{ \langle D_2(\omega^{k+1} x^k) \rangle
\over 1 - z(\omega^{k+1} x^k)}
= \sum_{k=1}^\infty k {\langle D_2(\omega^{k+1} x^k)\rangle
\over 1 - z(\omega^{k+1} x^k)}.
}

Combining \tone\ and \ttwo\ into \toplug\ gives the final result
\eqn\final{
\Z^{(1)}(x) = {\lambda \ln x \over 4 \pi^2}
\Z^{(0)}(x)\left[ \sum_{k=1}^\infty k { \langle  D_2(\omega^{k+1}
x^k)\rangle
\over 1- z(\omega^{k+1}
x^k)} + \sum_{k,m=1}^\infty \langle P D_2(\omega^{k+1} x^k, \omega^{m+1}
x^m)\rangle
\right],
}
as advertised already in \mainresult,
for the one-loop partition function of $\N = 4$ SYM theory on
$\R \times S^3$, expressed in terms of the free
partition function $\Z^{(0)}$, the elementary partition function
\zelem, and the traces \dtwotrace\ and \pdtwo.

\subsec{One-loop Hagedorn temperature}

The partition function $\Z(x)$ has a simple pole at the Hagedorn temperature
\eqn\aaa{
\Z(x) \sim {c \over x_{\rm H} - x},
}
where $c$ is an irrelevant overall numerical coefficient.
To compute the one-loop correction $\delta x_{\rm H}$ to the
Hagedorn temperature, we simply expand
\eqn\aaa{
{c \over x_{\rm H} + \delta x_{\rm H} - x}
= {c \over x_{\rm H} - x} \left[1
- { \delta x_{\rm H} \over x_{\rm H} - x} + \cdots \right].
}
When we compare this to \final\ and recall that $x_{\rm H}$ is
such that $z(x_{\rm H}) = 1$, we note that only the $k=1$ term 
in the sum of $\langle D_2(x^k) \rangle$ contributes to the double
pole at the Hagedorn temperature.\foot{It is possible that the
second term in \final\ develops a pole
at the Hagedorn temperature after evaluating the sum, in which case
it would add a finite correction to the one-loop Hagedorn temperature.
However, this does not happen in any of the subsectors that we studied,
and numerical evidence suggests that it does not occur here either.}
Reading off the residue of this pole, we find
\eqn\aaa{
\delta x_{\rm H} = - \lim_{x \to x_{\rm H}} \left[(
x_{\rm H} - x){\lambda \ln x \over
4 \pi^2}
{\langle D_2(x)\rangle \over 1 - z(x)}
\right] = - {2 \over 3} x_{\rm H} {\lambda \ln x_{\rm H} \over 4 \pi^2}
\langle D_2(x_{\rm H}) \rangle.
}
Remarkably, we find from \dtwotrace\ 
that $\langle D_2(x_{\rm H})\rangle = 3/4$, which gives
\eqn\aaa{
{\delta x_{\rm H} \over x_{\rm H}}
= - {\lambda \ln x_{\rm H} \over 8 \pi^2},
\qquad
{\rm and~hence} \qquad
{\delta T_{\rm H} \over T_{\rm H}}
= -{1 \over \ln x_{\rm H}} {\delta x_{\rm H} \over x_{\rm H}}
= {\lambda \over 8 \pi^2}.
}
The one-loop Hagedorn temperature is therefore
\eqn\aaa{
T_{\rm H} =
T_{{\rm H}}{(0)} \left( 1 + {\lambda \over 8 \pi^2} + \cdots\right), \qquad
T_{\rm H}{(0)} = {1 \over \ln(7 + 4 \sqrt{3})}.
}
It is encouraging that the sign is positive, consistent with the
simple guess that the Hagedorn temperature is a monotonically increasing,
smooth
function of $\lambda$ from $T_{\rm H} = T_0$ at zero coupling to
the AdS/CFT prediction that
$T_{\rm H} \sim \lambda^{1/4}$ at strong coupling.

\newsec{Discussion}

In this paper we have presented, in \mainresult,
the one-loop correction
to the partition function of $SU(N)$, $\N = 4$ SYM theory
on $\R \times S^3$ at infinite $N$ and below the Hagedorn temperature.

Several recent papers including
\refs{\AharonySX,\FidkowskiFC,\LiuVY,\AharonyIG}
address the thermodynamics and phase
transition structure of weakly coupled gauge theories with the goal of
smoothly connecting onto the strong coupling predictions implied
by the AdS/CFT correspondence.
String theory in $AdS_5 \times S^5$
has both a Hagedorn transition \WittenZW\ at $T_{\rm H} \sim \lambda^{1/4}$ and
a Hawking-Page
transition \HawkingDH\ involving the nucleation of $AdS_5$ black
holes at $T_{\rm HP} = 3/2 \pi$.
The authors of \AharonySX\ lay out the ways
in which the phase diagram of the weakly coupled theory
can be matched onto these strong coupling predictions.
Various qualitatively different phase diagrams can in part be
distinguished by the sign of a particular coefficient in 
the effective action for the Polyakov loop $U$, which is the
order parameter for the phase transition.
The computation of this coefficient
requires a three-loop calculation in thermal Yang-Mills
theory on $S^3$ which is in progress \ThreeLoop.

One of the motivations
for the present work was the desire to provide an independent
check of some pieces of the calculation of \ThreeLoop\ from
a completely orthogonal starting point.
The one-loop calculation in this paper is equivalent to a two-loop
calculation in thermal Yang-Mills theory and clearly has some, but not
complete,
overlap
with the work of \ThreeLoop. 
In one sense our calculation contains less information than
the effective action for the Polyakov loop because we integrate
out {\it all} degrees of freedom in the theory, including $U$.
Moreover, our method is only applicable for
temperatures below the Hagedorn transition.
On the other hand, our calculation contains more information since
we have also separately calculated, in \singletrace, the one-loop
correction to the partition function in the single-trace sector.
It requires extra work because the `trace' basis
it not particularly natural at finite temperature,
and the result is slightly
messy because of the appearance of `number theoretic' quantities,
such as the Euler function $\phi(n)$ or the condition $k:(k,L) = 1$
in \mainpolyaresult.  These always disappear at the end of the day
in any formula (such
as  \zfree\ or \final) which includes arbitrary-trace
operators.

The finer information present in the single-trace result \singletrace\ is
deeply related to  recent studies of integrability
in $\N = 4$ SYM theory because the one-loop correction to the
partition function is essentially just the trace of the one-loop
dilatation operator, and therefore encodes the sum of all anomalous
dimensions in the theory (sorted according
to bare dimension by the powers of
$x^{D_0}$).
The one essential subtlety
is that only cyclically invariant spin configurations
correspond to gauge theory operators, because traces of elementary
Yang-Mills fields are automatically cyclically invariant.
Although the spin chain Hamiltonian only acts on two neighboring
sites at a time,
the projection onto cyclically invariant states induces an effective
long-range interaction
on the spin chain which is irrelevant at temperatures near
the Hagedorn transition but dominates
the partition function, and significantly
complicates the calculation thereof,
at low temperatures.
Interestingly, the fact that the $PSL(4|4)$ Hamiltonian is integrable
\BeisertYB\ played absolutely no apparent role in our calculation.
It would be very interesting to understand our result in the
context of integrability.

It would also be interesting to consider higher-loop corrections
to our results.  Although the dilatation
operator of  the full $\N = 4$ theory is only known to one
loop, in the $SU(2)$ sector its precise form is known up to three
loops (at the planar level).
Furthermore, depending on the assumptions (such as integrability)
that one is willing to make, one can go all the way to five loops
\BeisertRY. 

\bigskip

\centerline{\bf Acknowledgments}

It is a pleasure to thank O.~Aharony,
S.~Minwalla, J.~Plefka, H.~Reall and R.~Roiban for useful
discussions, correspondence, and comments on the manuscript.
This research was supported in part by the National Science Foundation under
Grant No.~PHY99-07949.

\vfil

\appendix{A}{The Module $\V_1$}

We tabulate here the $SL(4) \times SL(2) \times SL(2)$
decomposition of the primary states in the module $\V_1$,
using the notation of \DolanZH\ (where it is referred
to as ${\cal{B}}^{ {1 \over 4}, {1 \over 4} }_{[1,0,1](0,0)}$).

$$
\hskip-1.0cm
\matrix{
\noalign{\vskip-6pt}
D_0\qquad\cr
2{\qquad}&
&&
{~~~}
&&
{~~~}
&&
{~~~}
&&
\hidewidth {\scriptstyle [1,0,1]_{(0,0)}}\hidewidth
&&
{~~~}
&&
{~~~}
&&
{~~~}
&&\cr
&&&&&&&&\Bsw
&{~~~~~~~~}&\Bse&&&&&&&\cr
{5\over 2}{\qquad}&&&&&&&\hidewidth {[1,0,0]_{({1 \over 2},0)}
\atop [0,1,1]_{({1\over 2},0)}}
\hidewidth&&&&
\hidewidth
{[0,0,1]_{(0,{1 \over 2})} \atop [1,1,0]_{(0,{1 \over 2})}}
\hidewidth&&&&&&\cr
&&&&&&\Bsw&&\Bse&&\Bsw&&\Bse&&&&&\cr
3{\qquad}&&&&&\hidewidth
{[0,1,0]_{(0,0)} \atop
[0,0,2],[0,1,0]_{(1,0)}}
\hidewidth &&&&
\hidewidth {[0,0,0], [0,2,0]_{({1 \over 2}, {1 \over 2})}
\atop  [1,0,1]_{({1 \over 2}, {1 \over 2})}}\hidewidth&&&&
\hidewidth
{[0,1,0]_{(0,0)} \atop
[2,0,0],[0,1,0]_{(0,1)}}
\hidewidth&&&&\cr
&&&&\Bsw&&\Bse&&\Bsw&&\Bse&&\Bsw&&\Bse&&&\cr
{\textstyle{7\over 2}}{\qquad}&&&\hidewidth
{\scriptstyle [0,0,1]_{({3 \over 2},0),({1 \over 2},0)}}
\hidewidth&&&& \hidewidth
{[0,0,1]_{({1 \over 2},1)} \atop
[1,1,0]_{({1 \over 2},1)}}\hidewidth &&&& \hidewidth
{[1,0,0]_{(1,{1 \over 2})} \atop
[0,1,1]_{(1,{1 \over 2})}}
\hidewidth &&&& \hidewidth
{\scriptstyle [1,0,0]_{(0,{3 \over 2}),(0,{1 \over 2})}}
\hidewidth&&\cr 
&&\Bsw&&\Bse &&\Bsw &&\Bse&&\Bsw&&\Bse&&\Bsw &&\Bse&\cr
4{\qquad}&\hidewidth
{\scriptstyle [0,0,0]_{(1,0)}}
\hidewidth&&&&\hidewidth
{\scriptstyle [0,1,0]_{({1 \over 2},
{3 \over 2})}}
\hidewidth &&&&
\hidewidth 
{[0,0,0],[1,0,1]_{(1,1)}, [0,2,0]_{(0,0)} \atop -[0,0,0],[1,0,1]_{(0,0)}}
 \hidewidth 
&&&&\hidewidth
{\scriptstyle [0,1,0]_{({3 \over 2},
{1 \over 2})}}
\hidewidth &&&&\hidewidth
{\scriptstyle [0,0,0]_{(0,1)}}\hidewidth\cr
&&&&&&\Bse &&\Bsw&&\Bse&&\Bsw &&&&&\cr
{\textstyle{9\over 2}}{\qquad}&&&&&&&\hidewidth 
{ [0,0,1]_{(1,{3 \over 2})} \atop
- [0,0,1],[1,1,0]_{(0,{1 \over 2})}}
\hidewidth&&&&
\hidewidth {[1,0,0]_{({3 \over 2},1)} \atop
-[1,0,0],[0,1,1]_{({1 \over 2},0)}}\hidewidth&&&&&&\cr
&&&&&&\Bsw &&\Bse&&\Bsw&&\Bse &&&&&\cr
5{\qquad}&&&&&\hidewidth
{\scriptstyle -[0,1,0]_{(0,1)}}\hidewidth &&&&
\hidewidth {[0,0,0]_{({3 \over 2},{3 \over 2})}
\atop -[0,0,0],[1,0,1]_{({1 \over 2},{1 \over 2})}}\hidewidth &&&&
\hidewidth
{\scriptstyle -[0,1,0]_{(1,0)}}\hidewidth &&&&\cr
&&&&&&\Bse &&\Bsw&&\Bse&&\Bsw &&&&&\cr
{\textstyle{11\over 2}}{\qquad}&&&&&&&\hidewidth
{\scriptstyle -[0,0,1]_{({1 \over 2},1)}}
\hidewidth&&&&
\hidewidth {\scriptstyle -[1,0,0]_{(1,{1 \over 2})}}
\hidewidth&&&&&&\cr
&&&&&&&&\Bse&&\Bsw&&&&&&&\cr
6{\qquad}&&&&&&&&&
\hidewidth {\scriptstyle -[0,0,0]_{(1,1)}}
\hidewidth &&&&&&&&\cr
}
$$

With the help of the
$SL(4) \times SL(2) \times SL(2)$ dimension formula
\eqn\aaa{
{\rm dim}[k,p,q]_{(j_1,j_2)}=
{1 \over 12}(k+p+q+3)(k+p+2)(p+q+2)(k+1)(p+1)(q+1)
(2 j_1 + 1)(2 j_2 + 1)
}
we can immediately read off the partition function for primary
states in $\V_1$,
\eqn\aaa{
V_1(x)
= 15 x^2 + 96 x^{5/2} + 252 x^3 + 336 x^{7/2} + 210 x^4 + 0 x^{9/2}
-84 x^5 - 48 x^{11/2} - 9 x^6,
}
in agreement with the result written in \vjdef.

\appendix{B}{Some Details}

In this appendix we show how to obtain the formula \rone\ 
from the matrix elements of $D_2$ given in \BeisertJJ.

\subsec{The $GL(4|4)$ oscillator basis}

The $GL(4|4)$ oscillator basis for $\A$ is realized by
a set of four bosonic oscillators ${\bf a}^\alpha$, ${\bf b}^{\dot{\alpha}}$
($\alpha,\dot{\alpha} \in \{1,2\}$)
and four
fermionic oscillators ${\bf c}^a$
($a \in \{1,2,3,4\}$) with the usual relations
\eqn\aaa{
[{\bf a}^\alpha, {\bf a}_\beta^\dagger]
= \delta^{\alpha}_{\beta}, \qquad
[{\bf b}^{\dot{\alpha}}, {\bf b}^\dagger_{\dot{\beta}}] =
\delta^{\dot{\alpha}}_{\dot{\beta}}, \qquad
\{ {\bf c}^a, {\bf c}^\dagger_b \} = \delta^a_b,
}
and a vacuum $|0\rangle$ annihilated by all of the lowering
operators.
The only constraint on physical states is that they should be annihilated
by the central charge
\eqn\aaa{
C = 1 - {1 \over 2} {\bf a}_\alpha^\dagger {\bf a}^\alpha
+ {1 \over 2} {\bf b}_{\dot{\alpha}}^\dagger
{\bf b}^{\dot{\alpha}} - {1 \over 2} {\bf c}_a^\dagger {\bf c}^a.
}
The tree-level dilatation operator corresponds to
\eqn\aaa{
D_0 = 1 + {1 \over 2} {\bf a}_\alpha^\dagger {\bf a}^\alpha
+ {1 \over 2} {\bf b}_{\dot{\alpha}}^\dagger
{\bf b}^{\dot{\alpha}}.
}

To consider two letters $\A \times \A$ we simply have two copies of
the above algebra, indexed by a subscript $(i) \in \{(1),(2)\}$.
A general state in $\A_{(i)}$ will be labeled by
its oscillator occupation numbers
$(a^1_{(i)},
a^2_{(i)},
b^{\dot{1}}_{(i)},
b^{\dot{2}}_{(i)},
c^1_{(i)},
c^2_{(i)},
c^3_{(i)},
c^4_{(i)}) \equiv N_{(i)}$.
In this basis we have
\eqn\need{\eqalign{
C_{(i)} &= 1 -
\half (a^1_{(i)} + a^2_{(i)})
+ \half (b^{\dot{1}}_{(i)} + b^{\dot{2}}_{(i)} )
- \half(c^1_{(i)}+c^2_{(i)}+c^3_{(i)}+c^4_{(i)}),\cr
D_{0(i)} &= 1 +
\half (a^1_{(i)} + a^2_{(i)})
+ \half (b^{\dot{1}}_{(i)} + b^{\dot{2}}_{(i)} ).
}}
Then to calculate a trace
$\Tr_{\A_{(i)}}$ literally means that we perform a sum
of the form
\eqn\tracea{
\sum_{\A_{(i)}} \equiv
\sum_{N_{(i)}} \delta(C_{(i)})=
\sum_{a^1_{(i)},
a^2_{(i)},
b^{\dot{1}}_{(i)},
b^{\dot{2}}_{(i)} = 0}^\infty
\sum_{c^1_{(i)},
c^2_{(i)},
c^3_{(i)},
c^4_{(i)} = 0}^1
\delta(C_{(i)})
}
over all possible oscillator numbers, subject to the
physical state constraint.
As a check, it is straightforward to confirm using this
formula and \need\ that
\eqn\aaa{
\Tr_\A[ x^{D_0}] = \sum_\A
x^{D_0}
= {2 x(3-\sqrt{x}) \over (1 - \sqrt{x})^3},
}
in agreement with \zelem.

Now we consider the action of the one-loop dilatation operator
$D_{2(12)}$ on two sites, following \BeisertJJ.
If we let ${\bf A}_I^\dagger =
({\bf a}_{\alpha}^\dagger, {\bf b}_{\dot{\alpha}}^\dagger,
{\bf c}_a^\dagger)$ schematically denote all of the raising operators,
then a general state in $\A \times \A$ can be written as
\eqn\aaa{
|s_1,\ldots,s_n;
\{ I_i \}\rangle =
{\bf A}^\dagger_{I_1(s_1)} \cdots {\bf A}^\dagger_{I_n(s_n)}|0\rangle,
}
where $s_i \in \{1,2\}$ indicates on which site the oscillator acts.
The dilatation operator does
not change the type or number of elementary $GL(4|4)$
oscillators, but can only cause them to hop from site 1
to 2 or vice versa according to the rule given in \BeisertJJ:
\eqn\dtwoelement{
D_{2(12)}
|s_1,\ldots,s_n; \{I_i\}\rangle
= \sum_{s_1',\ldots,s_n' = 1,2}
c(n,n_{12},n_{21})
\delta(C_{(1)}) \delta(C_{(2)})
|s_1',\ldots,s_n'; \{I_i\}\rangle,
}
where $n_{12},n_{21}$ 
count the number of oscillators hopping from site 1 to 2 or vice versa
and the coefficients $c(n,n_{12},n_{21})$ are
given in \coefficients.
In what follows we will consider a generalized operator of the
form \dtwoelement,
\eqn\qqdef{
Q|s_1,\ldots,s_n; \{I_i\}\rangle
= \sum_{s_1',\ldots,s_n' = 1,2}
q^n q_{12}^{n_{12}} q_{21}^{n_{21}}
|s_1',\ldots,s_n'; \{I_i\}\rangle,
}
with matrix elements $q^n q_{12}^{n_{12}} q_{21}^{n_{21}}$
instead of $c(n,n_{12},n_{21})$.

\subsec{The combinatorics of hopping}

We are therefore interested in the studying the combinatorics of
oscillators hopping between two sites.
Consider first a toy system with just a single bosonic
oscillator ${\bf a}$.  The most general state in $\A_{(1)}\times
\A_{(2)}$ would then be
\eqn\aaa{
|a_{(1)},a_{(2)}\rangle =
({\bf a}^\dagger_{(1)})^{a_{(1)}}
({\bf a}^\dagger_{(2)})^{a_{(2)}} |0\rangle
= \prod_{i=1}^{n} a^\dagger_{(s_i)} |0\rangle,
}
with
\eqn\aaa{
n = a_{(1)}+a_{(2)}, \qquad
s_i = \{  \underbrace{1,\ldots,1}_{a_{(1)}},
\underbrace{2,\ldots,2}_{a_{(2)}} \}.
}
A quantity of interest is the
matrix element
\eqn\aaa{
\eqalign{
h_{a_{(1)}, a_{(2)}}(q,q_{12},q_{21})&=
\langle a_{(1)},a_{(2)}| P Q |a_{(1)},a_{(2)}\rangle\cr
&=
\sum_{s_1',\ldots,s_n'=1}^2
q^n q_{12}^{n_{12}} q_{21}^{n_{21}}
\langle a_{(1)}, a_{(2)} | P 
\prod_{i=1}^{n} a^\dagger_{(s'_i)} |0\rangle,
}
}
where $Q$ is defined in \qqdef.
The function $h$ counts the number of ways that the initial state
$|a_{(1)}, a_{(2)}\rangle$ is mapped to itself, up to a permutation
$P$, under the action of an operator of the form
\dtwoelement, weighted according to the powers of $q_{12}$ and
$q_{21}$ which tell us the number of oscillators that have hopped
from 1 to 2 or vice versa.
An elementary combinatoric analysis reveals that
\eqn\oneboson{
\eqalign{
h_{a_{(1)}, a_{(2)}}(q,q_{12},q_{21})&=
\sum_{a=0}^\infty { a_{(1)} \choose a}
{a_{(2)} \choose a}
q^{a_{(1)} + a_{(2)}}
q_{12}^{a_{(1)} - a}
q_{21}^{a_{(2)} - a}\cr
&= (q q_{12})^{a_{(1)}}
(q q_{21})^{a_{(2)}}
F(-a_{(1)}, -a_{(2)}, 1, (q_{12} q_{21})^{-1}),
}}
where $F$ is the hypergeometric function.

A similar analysis can be done for the case of a single 
fermionic oscillator ${\bf c}$.  For a general
state $|c_{(1)},c_{(2)}\rangle$ we find
\eqn\onefermion{
g_{c_{(1)}, c_{(2)}}(q,q_{(12)},q_{(21})
= (q q_{12})^{c_{(1)}} (q q_{21})^{c_{(2)}}
\left(
1 - {c_{(1)} c_{(2)} \over q_{12} q_{21}}
\right).
}
Since fermionic oscillators only have occupation numbers
which are 0 or 1, this formula encodes the four cases
\eqn\aaa{
g_{0,0} = 1, \qquad
g_{0,1} = q q_{21}, \qquad
g_{1,0} = q q_{12}, \qquad
g_{1,1} = q^2 (q_{12} q_{21} - 1).
}
For $g_{0,0}$ there are no oscillators, so there is no hopping possible.
For $g_{0,1}$ we start with one oscillator on site 2, which moves
to site 1 giving a factor of $q_{21}$.
In the final case, $g_{1,1}$ we have one fermionic
oscillator on each site.  They can either stay where they
are, or they can flip, accounting for the two terms in $g_{1,1}$.

For a system with multiple oscillators we simply multiply
together the appropriate partition functions \oneboson\ and
\onefermion\ for the individual oscillators.
For $GL(4|4)$ this gives
\eqn\aaa{
\eqalign{
P_{N_{(1)}, N_{(2)}}(q,q_{12},q_{21})
&= \langle N_{(1)}, N_{(2)}|PQ |N_{(1)}, N_{(2)}\rangle
\cr
&=\prod_{\alpha=1}^2
h(a^\alpha_{(1)},a^\alpha_{(2)})
\prod_{\dot{\alpha}=1}^2
h(b^{\dot{\alpha}}_{(1)},b^{\dot{\alpha}}_{(2)})
\prod_{a=1}^4
g(c^a_{(1)}, c^a_{(2)}).
}}
The quantity
\eqn\combinatorics{
\eqalign{
R(w,y;q,q_{12},q_{21})&=
\Tr_{\A \times \A}
[ P w^{D_{0(1)}} y^{D_{0(2)}} Q]
\cr
&=
\sum_{N_{(1)},
N_{(2)}}
\delta(C_{(1)})
\delta(C_{(2)})
w^{D_{0(1)}}
y^{D_{0(2)}}
P_{N_{(1)}, N_{(2)}}(q,q_{12},q_{21})
}}
then represents a trace over $\A \times \A$ which
counts all
of the possible
hoppings between
a state $|N_{(1)},N_{(2)}\rangle$
and its permutation
$|N_{(2)},N_{(1)}\rangle$,
weighted appropriately by $w$ to the power of the dimension
of site 1, times $y$ to the power of the dimension of site 2,
times $q$ to the power of the total number of oscillators on
both sites, times $q_{12}$ to the power of the number of 
oscillators hopping from site 1 to 2, times $q_{21}$ to the
power of the number of oscillators hopping from site 2 to 1.
Concretely, we obtain from \tracea, \oneboson\ and
\onefermion\ the formula
\eqn\firstr{
\eqalign{
&R(w,y;q,q_{12},q_{21}) =
\sum_{a^1_{(1)},
a^2_{(1)},
b^{\dot{1}}_{(1)},
b^{\dot{2}}_{(1)},a^1_{(2)},
a^2_{(2)},
b^{\dot{1}}_{(2)},
b^{\dot{2}}_{(2)} = 0}^\infty
\sum_{c^1_{(1)},
c^2_{(1)},
c^3_{(1)},
c^4_{(1)},
c^1_{(2)},
c^2_{(2)},
c^3_{(2)},
c^4_{(2)} = 0}^1
\cr
&\qquad\qquad
\delta(C_{(1)})
\delta(C_{(2)})
(q q_{12})^{n_{(1)}} (q q_{21})^{n_{(2)}}
w^{D_{0(1)}} y^{D_{0(2)}}
\prod_{a=1}^4\left(
1 - {c^a_{(1)} c^a_{(2)} \over q_{12} q_{21}}
\right)\cr
&\qquad\qquad\qquad\times
\prod_{\alpha=1}^2
F(-a^\alpha_{(1)}, -a^\alpha_{(2)}, 1, (q_{12} q_{21})^{-1})
\prod_{\dot{\alpha}=1}^2
F(-b^{\dot{\alpha}}_{(1)}, -b^{\dot{\alpha}}_{(2)}, 1, (q_{12} q_{21})^{-1})
}}
where
\eqn\aaa{
n_{(i)} = \sum_{\alpha=1}^2
a_{(i)}^\alpha +
\sum_{\dot{\alpha}=1}^2 b_{(i)}^{\dot{\alpha}}
+ \sum_{a=1}^4 c^a_{(i)}
}
denotes the total number of oscillators on site $i$.

The first step in simplifying
\firstr\ is to rearrange the bosonic sums according to
\eqn\bsum{
\sum_{a_{(i)}^1, a_{(i)}^2 = 0}^\infty f(a_{(i)}^1, a_{(i)}^2)
= \sum_{s_{(i)} = 0}^\infty
\sum_{t_{(i)}=0}^{s_{(i)}}
f(t_{(i)}, s_{(i)}- t_{(i)}).
}
(For the $b$ oscillators we will use
$\dot{s}$ and $\dot{t}$ as the new summation variables.)
After this substitution, the $t$ variables only appear
in the arguments of the hypergeometric functions.
The sum over $t_{(1)}$ and $t_{(2)}$ can be done with the help of
the identity
\eqn\aaa{\eqalign{
&\sum_{t_{(1)}=0}^{s_{(1)}}
\sum_{t_{(2)}=0}^{s_{(2)}}
F(-t_{(1)},-t_{(2)},1,z)
F(-s_{(1)}+t_{(1)},-s_{(2)}+t_{(2)},1,z)
\cr
&\qquad\qquad\qquad\qquad\qquad\qquad\qquad\qquad= (1+s_{(1)})
(1 + s_{(2)}) F(-s_{(1)}, -s_{(2)},2,z).
}}

The sum over fermionic occupation numbers can be similarly simplified
with the formula
\eqn\fsum{
\eqalign{
&\sum_{c^1_{(1)},
c^2_{(1)},
c^3_{(1)},
c^4_{(1)},
c^1_{(2)},
c^2_{(2)},
c^3_{(2)},
c^4_{(2)} = 0}^1
f\left(\sum_{a=1}^4 c^a_{(1)},
\sum_{a=1}^4 c^a_{(2)}\right)
\prod_{a=1}^4 (
1 - {c^a_{(1)} c^a_{(2)}} z)
\cr
&\qquad\qquad\qquad=
\sum_{F_{(1)}=0}^4 \sum_{F_{(2)}=0}^4 f(F_{(1)},F_{(2)}) \sum_{j=0}^4
(-1)^j z^j {4 \choose j} {4 - j \choose F_{(1)} - j} 
{4 - j \choose F_{(2)} - j}.
}}
The results \bsum\ and \fsum\ allow \firstr\ to be written as
\eqn\secondr{
\eqalign{
R(w,y;q,q_{12},q_{21})&=
\sum_{s_{(1)},
s_{(2)},
\dot{s}_{(1)},
\dot{s}_{(2)}=0}^\infty
\sum_{F_{(1)},F_{(2)},j=0}^4
(-1)^j z^j {4 \choose j}
{4-j \choose F_{(1)} - j}
{4 - j \choose F_{(2)} - j}
\cr
&\qquad\times
w^{1 + s_{(1)}/2 + \dot{s}_{(1)}/2}
y^{1 + s_{(2)}/2 + \dot{s}_{(2)}/2}
\cr
&\qquad \times (q q_{12})^{n_{(1)}}
(q q_{21})^{n_{(2)}}
 F(-s_{(1)}, -s_{(2)}, 2, z)
F(-\dot{s}_{(1)}, -\dot{s}_{(2)}, 2, z)\cr
&\qquad \times
\prod_{i=1}^2
\delta(1 - \half s_{(i)}+ \half
\dot{s}_{(i)} - \half F_{(i)})
(1+s_{(i)})(1 + \dot{s}_{(i)}),
}}
where
\eqn\ni{
z = {1 \over q_{12} q_{21}}, \qquad
n_{(i)} = s_{(i)} + \dot{s}_{(i)} + F_{(i)}.
}

Of course, we are not particularly interested in the quantity
$R(w,y;q,q_{12},q_{21})$.  It is simply an auxiliary quantity
which counts the possible ways for the operators to hop between
sites.  We are interested in the trace of \dtwoelement\ instead
of the trace of \qqdef, which we can calculate via the replacement
\eqn\subs{
\langle P D_2(w,y)\rangle
=\left. R(w,y;q,q_{12},q_{21}) \right|_{q^n q_{12}^{n_{12}}
 q_{21}^{n_{21}} \to
c(n,n_{12},n_{21})},
}
where the notation means simply that we expand \secondr\ in
powers of $q$, $q_{12}$ and $q_{21}$ and then make the substitution
indicated for the various powers.

In order to proceed, we expose the powers of $q_{12}$ and $q_{21}$
in \secondr\ by using the identity
\eqn\regf{
\eqalign{
&F(-s_{(1)},-s_{(2)},2,z) F(-\dot{s}_{(1)},-\dot{s}_{(2)},2,z)
\cr
&
= \sum_{k=0}^\infty
z^k {\dot{s}_{(1)}! \dot{s}_{(2)}! \over k! (k+1)!}
{\cal{F}}(-1-k,-k,-s_{(1)},-s_{(2)};2,1-k+\dot{s}_{(1)},1-k+\dot{s}_{(2)};1)
}}
where ${\cal{F}}$ is the regularized hypergeometric function.
Combining \subs, \regf\ and \secondr\ finally leads to the
formula \rone\ for the desired trace, after some simplification
of the notation.

\listrefs
\end